\documentclass[aps,pra,showpacs,superscriptaddress,groupedaddress, twocolumn,nofootinbib]{revtex4-2}  

\usepackage[utf8]{inputenc}
\usepackage{bm}
\usepackage{color}

\usepackage{hyperref}
\usepackage{amsmath, mathtools}
\usepackage{amssymb}
\usepackage{wasysym}
\usepackage{graphicx} 
\usepackage{booktabs} 
\usepackage{pdfsync}
\usepackage{verbatim}
\usepackage{latexsym}
\usepackage{dsfont}

\usepackage{float}

\usepackage[bottom]{footmisc}
\usepackage{subcaption}

\def\sec#1{\section{#1} }
\def\ssec#1{\subsection{#1} }

\def\({\left(}
\def\){\right)}
\def\[{\left[}
\def\]{\right]}

\def\a{\alpha}
\def\b{\beta}
\def\f#1#2{\frac{#1}{#2}}
\def\g{\gamma}

\def\d{\partial}
\def\de{\delta}

\def\del{\nabla}

\def\vep{\varepsilon}

\def\l{\lambda}
\def\L{\Lambda}
\def\m{\mu}
\def\n{\nu}

\def\p{\pi}

\def\s{\sigma}

\def\<{\langle}
\def\>{\rangle}

\providecommand{\abs}[1]{\left\lvert#1\right\rvert}

\usepackage{graphicx}
\usepackage{dcolumn}
\usepackage{bm}

\begin{document}
\title{Relativistic Ritz approach to hydrogen-like atoms I: theoretical considerations}

\author{David~M.~Jacobs}
\email{djacobs@norwich.edu}
\affiliation{Physics Department, Norwich University, 158 Harmon Dr, Northfield, VT 05663, USA}
\affiliation{CERCA, Physics Department,
Case Western Reserve University,
Cleveland, OH 44106, USA}

\date{\today}


\begin{abstract}
The Rydberg formula along with the Ritz quantum defect ansatz has been a standard theoretical tool used in atomic physics since before the advent of quantum mechanics, yet this approach has remained limited by its non-relativistic foundation. Here I present a long-distance \emph{relativistic} effective theory describing hydrogen-like systems with arbitrary mass ratios, thereby extending the canonical Ritz-like approach. Fitting the relativistic theory to the hydrogen energy levels predicted by bound-state QED indicates that it is superior to the canonical, nonrelativistic approach. An analytic analysis reveals nonlinear consistency relations within the bound-state QED level predictions that relate higher-order corrections to those at lower order, providing guideposts for future perturbative calculations as well as insights into the asymptotic behavior of Bethe logarithms. Applications of the approach include fitting to atomic spectroscopic data, allowing for the determination the fine-structure constant from large spectral data sets and also to check for internal consistency of the data independently from bound-state QED.
\end{abstract}
\maketitle

\sec{Introduction}\label{Sec:intro}

The Standard Model of particle physics is an impressive scientific achievement, but is unlikely to be the final chapter in our quest to describe nature at a fundamental level. Several outstanding issues have been identified that appear to require physics beyond the Standard Model, including the observed matter/anti-matter asymmetry of the Universe \cite{Dine:2003ax} and the many observations supporting the dark matter hypothesis \cite{Bertone:2016nfn}. Accelerator laboratories have been used for nearly a century to perform particle physics experiments at increasingly higher energies, pushing the limits of our knowledge. However, in recent decades it has also become more feasible to seek new frontiers through the precision study of atoms and molecules \cite{safronova2018search}.

Within the Standard Model, quantum electrodynamics (QED) provides a description of the interaction between charged electrically-charged particles and photons with an accuracy unrivaled by anything in the physical sciences. The most striking agreement is arguably between the measurement of the electron g-factor \cite{Hanneke:2008tm, Hanneke:2010au} and theoretical predictions (see, e.g., \cite{Aoyama:2019ryr}) using the fine-structure constant ($\a$) as an experimental input \cite{Parker:2018vye,Morel:2020dww}, agreeing at a level below 1 part per trillion. It remains imperative to continue to refine such empirical comparisons not only to see if any improvement in the theory is needed, but also to probe for possible new physics beyond the Standard Model.

Improving the precision of the theoretical description of atoms within bound-state quantum electrodynamics (BSQED) is increasingly difficult at higher orders in perturbation theory (see, e.g., \cite{Indelicato_2019}) and, at the same time, spectroscopic experiments can be subject to unanticipated sources of error \cite{Khabarova:2021trt}. Having an effective theory of atoms and molecules may be of use in guiding future progress on both theoretical and experimental fronts.  The idea behind an effective theory is simple but profound: nature has for many physical systems allowed us to describe phenomena at long-distances without explicitly modeling the shorter-ranged interactions. Such an approach is ubiquitous in the physical sciences and effective field theory, in particular, has had much contemporary use in condensed matter, particle physics and cosmology; see, e.g., \cite{burgess2020introduction}.

Here I apply an effective approach to the quantum mechanics of hydrogen-like atoms. Consider that no atomic system is purely Coulombic; the effects of finite particle size and spin coupling depend on the system in question, but field-theoretic effects like vacuum polarization and self energy corrections are always present.  However, for single-electron atoms, all of these effects have a characteristic length scale\footnote{Here the Heaviside-Lorentz unit choice $\hbar=c=\vep_0=1$ is used except where otherwise noted.} less than or equal to the (reduced) Compton wavelength of the electron, $r_\text{\tiny QED}=m_e^{-1}$.  Some interactions decay with distance exponentially and are effectively local. Others are nonlocal, decaying only as a power law, but faster than the leading (scale-free) Coulomb term. Consider the common case in which an atom is composed of a nuclear charge $Ze$ and is bound with a particle of charge $-e$.  The characteristic atomic orbital radius is $r_\text{atom}=n^2/(m_\text{red}Z\a)$, where $m_\text{red}$ is the reduced mass, $\a=e^2/(4\pi)$ is the fine-structure constant, and $n$ is the integer principal quantum number. It is hence plausible for a long-range effective theory to have validity if the ratio of these length scales,
\begin{equation}
\frac{r_\text{\tiny QED}}{r_\text{atom}}=\frac{m_\text{red}}{m_e}\frac{Z\a}{n^2}\,,
\end{equation}
is small. Alternatively, one may consider this as the ratio of two energy scales,
\begin{equation}
\frac{E_\text{\tiny atom}}{\Lambda}=\frac{m_\text{red}}{m_e}\frac{Z\a}{n^2}\,,
\end{equation}
where, $E_\text{\tiny atom}=m_\text{red} \(Z\a\)^2/n^2$ and $\Lambda=m_eZ\a$. In any case, for electronic hydrogen, deuterium, and positronium this is ratio is small even for the ground state, $n=1$. For other hydrogen-like systems, such as muonic hydrogen or multi-electron atoms, we should expect that $n\gg 1$ may be necessary to make reliable predictions.

\begin{figure}[h!tb]
  \begin{center}
    \includegraphics[scale=.4]{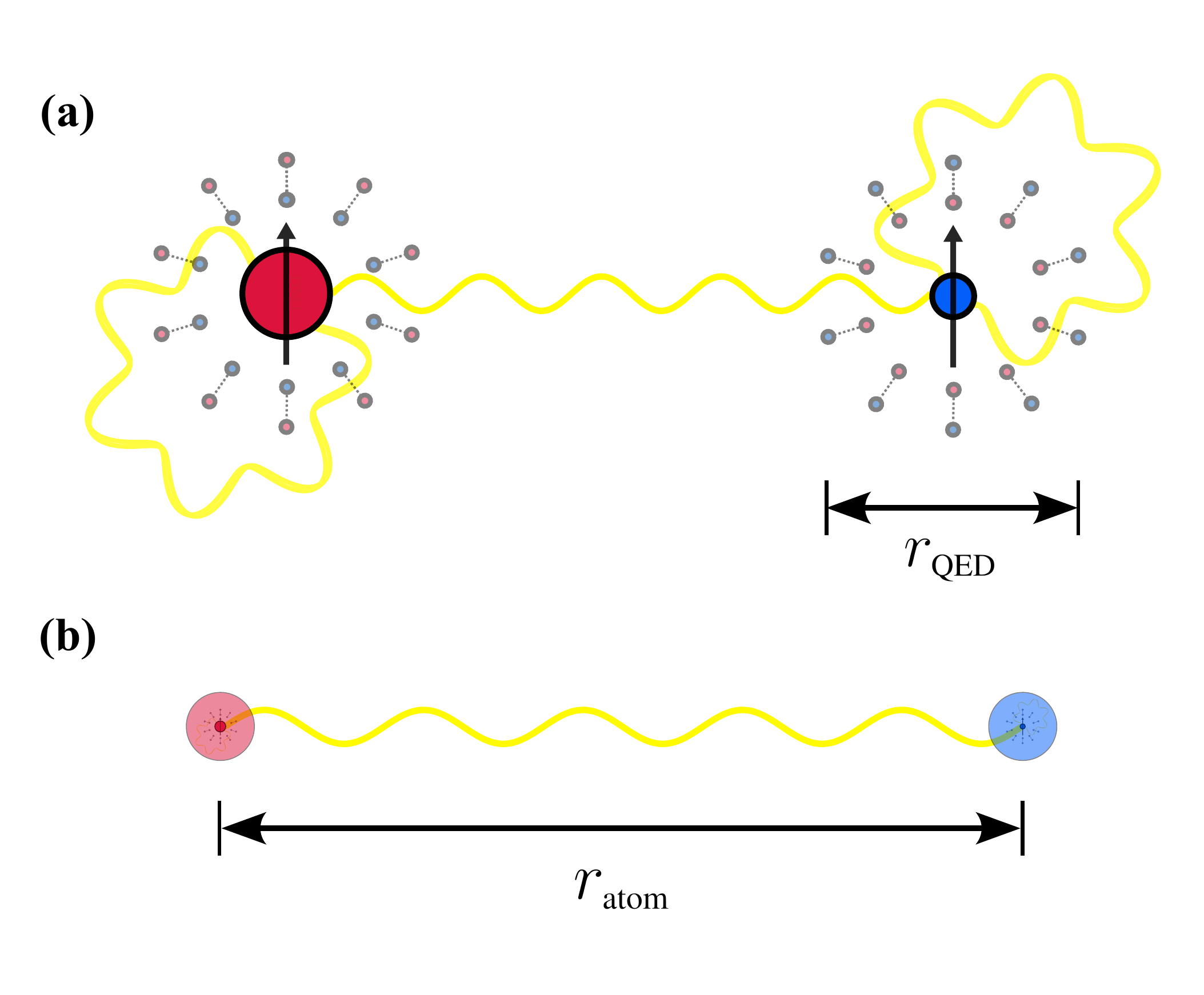}
  \end{center}
  \caption{\raggedright {\bf Hydrogen-like atom illustration.} {\bf (a)} Two generic charged particles bound electromagnetically. The QED length scale, $r_\text{\tiny QED}=\tfrac{\hbar}{m_e c}$, associated with short-ranged effects, such as spin-coupling, vacuum polarization, and self-energy corrections. {\bf (b)} Charged particles within the hydrogen atom exchanging a single Coulomb-like virtual photon. Even in the ground state, the characteristic atomic separation of the proton and electron, $r_\text{atom}=\tfrac{r_\text{\tiny QED}}{\a}$ is much bigger than $r_\text{\tiny QED}$, meaning a long-distance effective theory may be possible.}
  \label{quantum_fuzz_simplified}
  \end{figure}

Quantum defect theory \cite{Seaton_1983} would be the most obvious thing to try first as an effective theory in this context; it has been vital for the description of Rydberg states of multi-electron atoms \cite{gallagher_1994}. However, that effective theory is limited by its nonrelativistic foundation. With that approach, a two-particle system with masses $m_1$ and $m_2$ has energy levels parameterized by the Rydberg-Ritz formula. Within the modern quantum mechanical framework, this formula is 
\begin{equation}\label{NR_Ritz}
E_\text{Ryd-Ritz}=-\f{1}{2}\f{m_1 m_2}{m_1+m_2} \(\f{Z\a}{n_\star}\)^2\,,
\end{equation}
where the effective quantum number $n_\star=n-\de_n$ and the quantum defect, $\de_n$, effectively parametrizes deviations from the Coulomb potential at short distance. It is generally assumed to obey a series expansion in energy, such as the extended Ritz formula
\begin{equation}\label{extended_NR_Ritz}
\de_n=\de_{(0)}+ \f{\de_{(2)} }{\(n-\de_n\)^2 }+\f{ \de_{(4)} }{\(n-\de_n\)^4}  +\dots
\end{equation}
which extends the early analysis performed by Ritz \cite{ritz1908new} and is justified through a nonrelativistic quantum mechanical analysis  \cite{hartree_1928}. Alternatively, the modified Ritz formula,
\begin{equation}\label{modified_NR_Ritz}
\de_n=\de_{(0)}+ \f{\de_{(2)} }{\(n-\de_{(0)}\)^2 }+\f{ \de_{(4)} }{\(n-\de_{(0)}\)^4}  +\dots
\end{equation}
is commonly used.  However, equation \ref{NR_Ritz} cannot properly account for all relativistic effects, as discussed in Ref. \cite{Jacobs:2021cdb}.  This is especially problematic for hydrogen-like atoms wherein kinetic relativistic corrections are of the same order of magnitude as the shorter-ranged relativistic corrections, such as spin-orbit coupling. 
 
 To demonstrate the limitations of this standard \emph{nonrelativistic Ritz} approach, consider an infinitely massive ($m_2\to \infty$) and spinless nucleus with charge $Ze$ and zero size\footnote{This analysis is complementary to that found in Ref. \cite{Jacobs:2021cdb}, but I thank Gordon Drake for prompting me to consider it.}. If all field theoretic effects are ignored, the energy levels are exactly calculable as eigenvalues of Dirac-Coulomb equation,
\begin{equation}\label{canonical_Dirac-Coulomb}
E_\text{Dirac}=m_1\(\f{1}{\sqrt{    1+\(\f{Z\a}{n-\de_\text{Dirac}}\)^2   }          } -1\)\,,
\end{equation}
where the $n$-independent constant
\begin{equation}\label{canonical_Dirac-Coulomb_defect}
\de_\text{Dirac}=\(j+1/2\)-\sqrt{\(j+1/2\)^2-\(Z\alpha\)^2}
\end{equation}
depends on the electron's total angular momentum, $j$.

If equation \eqref{NR_Ritz} is capable of reproducing the energy levels given by \eqref{canonical_Dirac-Coulomb}, we should be able to determine the values of $\de_{(i)}$ necessary to do so.  Assuming $Z\a<1$, it follows also that $\de_\text{Dirac}<1$ and thus matching the expansions of both formulae in small $n^{-1}$ order by order in powers of $n^{-1}$ should accomplish this. Asymptotically,
\begin{equation}
\f{E_\text{Ryd-Ritz}}{m_1\(Z\a\)^2} = -\f{1}{2n^2}  -\f{\de_{(0)}}{n^3} -\f{3}{2}\f{\de_{(0)}^2}{n^4}+{\cal O}\(n^{-5}\)\,,
\end{equation}
whereas
\begin{equation}\label{Dirac_Expansion}
\f{E_\text{Dirac}}{m_1\(Z\a\)^2} = -\f{1}{2n^2}  -\f{\de_\text{Dirac}}{n^3} -\f{3}{2}\f{\de_\text{Dirac}^2}{n^4} +\f{3}{8}\f{\(Z\a\)^2}{n^4}+{\cal O}\(n^{-5}\)\,.
\end{equation}
Matching at order $n^{-3}$, it follows that
\begin{equation}
\de_{(0)}=\de_\text{Dirac}\,.
\end{equation}
However, at order $n^{-4}$ the matching cannot be completed because of the fourth term in equation \eqref{Dirac_Expansion}, and it cannot be accounted for without a very specific modification to either \eqref{extended_NR_Ritz} or \eqref{modified_NR_Ritz} that does not seem to be generalizable to other systems. This offending term has nothing to do with spin-orbit coupling, which falls off with distance faster than the leading $1/r$ Coulomb interaction, but is a motional (or kinetic) relativistic correction that is present for all particles, regardless of spin. Given that it has been standard practice (see, e.g., Ref. \cite{Kramida:2010lii}) to apply the nonrelativistic Ritz approach of formula \eqref{NR_Ritz} to fits of atomic spectral data\footnote{As noted in \cite{Kramida:2010lii}, a polarization formula may used as an alternative to \eqref{NR_Ritz}, at least for the purpose of extracting an ionization energy from spectral data. However, such a formula neglects higher order relativistic and relativistic recoil effects.}, it is even more pertinent to revisit this issue.

Attempts to construct a relativistic quantum defect theory for the Dirac-Coulomb system may be found in the literature, e.g. Ref. \cite{Johnson_1979}, but such analyses are appropriate only to the extent that the finite nuclear mass, i.e. recoil effects, are ignored.  In Ref. \cite{Jacobs:2021xgp} a relativistic quantum defect theory was derived for positronium, the special case in which $m_1=m_2$; generalizing that work to systems with arbitrary mass ratios is non-trivial and constitutes Section \ref{Sec:Model_Summary} of the present article. In Section \ref{QED_level_match} a matching between this relativistic effective theory and the numerical BSQED energy level predictions demonstrates its accuracy and superiority over the nonrelativistic Ritz approach. In Section \ref{QED_insights} an asymptotic analytic matching between those energy levels reveals non-linear consistency relations within BSQED that appear to apply to all hydrogen-like systems. In Section \ref{Bethe_Log_Discussion} implications for the Bethe logarithm are described. A brief conclusion is presented in Section \ref{Sec:Discussion}.

\sec{Model Derivation}\label{Sec:Model_Summary}

\ssec{Matching Calculation}
Following the procedure first explored in \cite{Jacobs:2021xgp}, let us seek an effective time-independent Schrodinger equation that describes the long distance interaction of two charged particles, namely
\begin{equation}\label{SchrodingerEqn}
\(\vep_1-m_1  + \vep_2-m_2 + U_\text{eff}\(r\)\)\psi=E\psi\,,
\end{equation}
where $r$ is the relative separation of the two particles, $\vec{p}\equiv\vec{p}_1=-\vec{p}_2$ is the conjugate momentum operator for the center-of-momentum motion, and $E$ is the energy of the system less the two masses. The relativistic energy operators
\begin{equation}
\vep_i=\sqrt{\vec{p}^{\,2}+m_i^2}\,,
\end{equation}
could, in principle, be implemented by expanding to any desired order in the quantities $\vec{p}^{\,2}/m_i^2$. It should be understood that $\vep_i$ represents an operator when working in configuration space, whereas it is a c-number in Fourier space.

By matching the calculation of the differential scattering cross section of the two particles using (a) the field-theoretical method and (b) the Born approximation to scattering with \eqref{SchrodingerEqn}, we may obtain the effective potential, $U_\text{eff}(r)$. Following \cite{berestetskii1982quantum}, the field theoretic calculation of the differential elastic scattering cross section yields
\begin{equation}\label{diffcross_QFT}
\f{d\s}{d\Omega}= \f{1}{64\p^2 \(\vep_1+\vep_2\)^2} \abs{M_{fi}}^2 \,.
\end{equation}
Here we compute the scattering amplitude, $M_{fi}$, for two distinguishable spin-$\tfrac{1}{2}$ particles and discover below that the result is the same for scalar particles. Take the electric charges to be $e_1$ and $e_2$, assuming that they scatter elastically with momentum transfer
\begin{equation}
\vec{q}\equiv\vec{p}_1\,'-\vec{p}_1.
\end{equation}
The lowest order QED amplitude, due to a single photon exchange, is
\begin{equation}\label{lowest_order_amplitude}
M_{fi}=e_1e_2\(\bar{u}_1'\g^\m u_1\)D_{\m \n}(q)\(\bar{u}_2'\g^\n u_2\)\,,
\end{equation}
where $\bar{u}_i'=\bar{u}(\vec{p}_i\,',m_i)$. The photon propagator in the Feynman (standard) gauge is
\begin{equation}
D_{\m \n}(q)=\f{4\p}{q^{2}} g_{\m\n}\,,
\end{equation}
and the square of the virtual photon four-momentum,
\begin{equation}
q^{2}=\omega^2-\vec{q}^{\,2}\,.
\end{equation}
 As the $\vec{q}\to 0$ limit is taken,
\begin{equation}\label{spin_series_form}
\bar{u}_i'\g^\m u_i   = 2p^\m_i + {\cal O}\(\abs{\vec{q}}\)\,, 
\end{equation}
which, at leading order, neglects any spin-dependent effects. Temporarily reintroducing factors of $\hbar$, loop diagrams that give relative corrections to \eqref{lowest_order_amplitude} must, for dimensional reasons, scale as some positive power of $\tfrac{\hbar \abs{\vec{q}}}{E_\text{light}}$, i.e.,
\begin{equation}\label{qtozero_matrix_element}
M_{fi} = \f{16\pi e_1 e_2}{q^2} \( \vep_1\vep_2  +\vec{p}^{\,2} \)\(1+{\cal O}\(\f{\hbar\abs{\vec{q}}}{E_\text{light}}\) \)   \,,
\end{equation}
where $E_\text{light}$ is the smallest relevant energy scale. In the context of the Standard Model, $E_\text{light}$ is no less than $m_e$. In dropping the correction terms in \eqref{qtozero_matrix_element}, we omit all spin-dependent and loop corrections that contribute to fine and hyperfine structure. All such terms diminish in the real-space potential with distance faster than $r^{-1}$. To re-iterate, QED is only assumed to be valid and sufficient for very low momentum-exchange scattering, i.e. at very large spatial separation.%

Omission of the terms described above has a very important simplifying consequence for the analyses that follow. Two propositions are made which we confirm below. First, we assume $U_\text{eff}(r) \propto r^{-1}$ at leading order, which is plausibly true given equation \eqref{qtozero_matrix_element}. Second, we suppose that \eqref{SchrodingerEqn} will yield an equation that displays the long distance behavior
\begin{equation}\label{psqr_posit}
 \lim_{r\to\infty} \vec{p}^{\,2} \psi \sim \(C_0+\f{C_1}{r} \)\psi\,,
\end{equation}
for some c-numbers, $C_{0}$ and $C_{1}$. This means that the commutation relation
\begin{equation}\label{commutation_simplification}
 \lim_{r\to\infty} \Big[\vec{p}^{\,2},U_\text{eff}(r)\Big]\psi\sim0~~~~\mbox{(effectively)}\,,
\end{equation}
may be used because the terms omitted in \eqref{commutation_simplification} decay with distance faster than $r^{-2} \psi$ and are therefore subdominant at long distance to each term in the commutator.

To use \eqref{SchrodingerEqn} for the analysis of a scattering event in the long-distance limit, we assume the two incoming particles to have momenta $\vec{k}$ and $-\vec{k}$, and make the identifications
\begin{eqnarray}
E&\equiv&\vep_1+\vep_2 - m_1 - m_2\notag\\
\vep_{i}&\equiv&\sqrt{k^2+m_{i}^2}\,.
\end{eqnarray}
Equation \eqref{commutation_simplification} allows us to use purely algebraic means to solve \eqref{SchrodingerEqn} for $\vec{p}^{\,2}\psi$:
\begin{equation}
\vec{p}^{\,2} \psi = \[k^2  - \f{2\vep_1 \vep_2}{\vep_1+\vep_2}U_\text{eff}(r) + {\cal O}\(U_\text{eff}^2\) \]\psi\,.
\end{equation}
A first-order scattering analysis then yields
\begin{equation}\label{diffcross_Schrodinger}
\f{d\s}{d\Omega}=  \f{1}{4\p^2}\(\f{\vep_1\vep_2}{\vep_1+\vep_2}\)^2 \abs{\tilde{U}_\text{eff}(\vec{q})}^2\,,
\end{equation}
where $\tilde{U}_\text{eff}(\vec{q})$ is the Fourier transform of $U_\text{eff}(r)$, and the factor $\vep_1 \vep_2/(\vep_1+\vep_2)$ can be interpreted as a relativistic generalization of the reduced mass.%

Finally, by using \eqref{qtozero_matrix_element} to match  \eqref{diffcross_QFT} with \eqref{diffcross_Schrodinger}, we find the effective potential in Fourier space to be 
\begin{equation}\label{U_gen_q-space}
\tilde{U}_\text{eff}(\vec{q})=-\f{4\pi e_1e_2}{q^2}\(1+\f{\vec{p}^{\,2}}{\vep_1\vep_2}   \)\,,
\end{equation}
where this choice of sign is consistent with the known nonrelativistic behavior of the electric potential energy between two charged particles.

\ssec{Real Space Effective Potential and Solutions}

Let us now specialize to the case of $e_1=-e$, and $e_2=+Ze$ for some positive integer, $Z$. The $q^2$ in the denominator of \eqref{U_gen_q-space} makes a the Fourier transform to real space challenging; this complication is simply not present in the special equal-mass case considered in \cite{Jacobs:2021xgp}. To proceed 
 note that, given the definition
\begin{equation*}
\vec{q}\equiv \vec{p}_1\,'-\vec{p}_1\,,
\end{equation*}
the energy of the virtual photon obeys 
\begin{equation}\label{omega_eqn}
\lim_{\vec{q}\to0}\omega=\f{\vec{p}\cdot\vec{q}}{2\vep_1 \vep_2}\(\vep_2-\vep_1\)\,.
\end{equation}
In this limit, its squared four-momentum is
\begin{equation}
q^2= \(\vec{p}\cdot\vec{q}\)^2\(\f{\(\vep_2-\vep_1\)}{2\vep_1 \vep_2}\)^2 - \vec{q}^{\,2}\,,
\end{equation}
and from \eqref{U_gen_q-space} it follows that
\begin{equation}\label{U_eff_q-space_01}
\tilde{U}_\text{eff}(\vec{q})=-\f{4\pi Z\alpha}{\vec{q}^{\,2}}\f{\(1+\f{\vec{p}^{\,2}}{\vep_1\vep_2}   \)}{1 - \f{\(\vec{p}\cdot\vec{q}\)^2}{\vec{q}^{\,2}}\(\f{\(\vep_2-\vep_1\)}{2\vep_1 \vep_2}\)^2}\,.
\end{equation}
In Fourier space the following inequality is obeyed
\begin{equation}
\f{\(\vec{p}\cdot\vec{q}\)^2}{\vec{q}^{\,2}}\(\f{\(\vep_2-\vep_1\)}{2\vep_1 \vep_2}\)^2 \leq \frac{1}{4}\,,
\end{equation}
which means that \eqref{U_eff_q-space_01} may be written as a geometric series
\begin{multline}\label{U_eff_q-space_02}
\tilde{U}_\text{eff}(\vec{q})=-\f{4\pi Z\alpha}{\vec{q}^{\,2}}                        \(1+\f{\vec{p}^{\,2}}{\vep_1\vep_2}   \)\\
\times   \sum_{n=0}   \(\f{\(\vec{p}\cdot\vec{q}\)^2}{\vec{q}^{\,2}}\(\f{\(\vep_2-\vep_1\)}{2\vep_1 \vep_2}\)^2\)^n\,.
\end{multline}
For the special case in which $m_1=m_2$, considered in \cite{Jacobs:2021xgp}, only the $n=0$ term is non-zero and gives a Coulomb-like result. For the general case all terms must be considered, so the real-space effective potential has the form
\begin{equation}
U_\text{eff}(r)\propto \sum_{n=0} I^{(n)}\,,
\end{equation}
where
\begin{equation}\label{In_integral_def}
I^{(n)}=\int \f{d^3q}{\(2\pi\)^3} e^{i\vec{q}\cdot\vec{r}} \f{1}{\vec{q}^{\,2}} \(\f{\(\vec{p}\cdot\vec{q}\)^2}{\vec{q}^{\,2}}\)^n\,.
\end{equation}
For $n=0$,
\begin{equation}
I^{(0)}=\!\frac{1}{4\pi r}\,,
\end{equation}
and for $n=1$,
\begin{equation}\label{n=1_Fourier}
I^{(1)}=\! \frac{1}{8\pi r} \(\vec{p}^{\,2}  -\f{\vec{r}\cdot\(\vec{r}\cdot\vec{p}\)\vec{p}}{r^2}  \) \,,
\end{equation}
a result found readily in the literature, e.g.,  \cite{berestetskii1982quantum}.  Consider that $\vec{p}^{\,2}=-\d_r^2-\f{2}{r}\d_r+\f{1}{r^2}\del_\Omega^2$ and $\vec{r}\cdot\(\vec{r}\cdot\vec{p}\)\vec{p}=-r^2 \d_r^2$. This means that the $n=1$ term acting on $\psi$ scales as
\begin{equation}\label{operator_scaling}
{\cal O}\(\f{1}{r^2}\d_r \psi\)~~~~\text{or}~~~~{\cal O}\(\f{1}{r^3} \psi\)\,,
\end{equation}
which, at large distance, is negligibly small compared to the Coulomb-like $n=0$ term.

For arbitrary integer $n$, discussed below, a different tack is taken, but the conclusion is that all $n\neq0$ terms acting on $\psi$ fall off with distance faster than $r^{-1}\psi$ and therefore may be discarded when compared to the leading Coulomb-like term. Choosing coordinates such that $\vec{r}$ is aligned with the $z$-axis so that $\vec{q}\cdot\vec{r}=q r \cos{\theta}$,  let $w=qr$ and $x=\cos{\theta}$, integrate over the azimuthal angle, and write equation \eqref{In_integral_def} as
\begin{equation}\label{alt_In_def}
I^{(n)}=\f{1}{4\p^2}\f{1}{r}\int_0^\infty dw  \int_{-1}^1 dx\, e^{i w x} \(\hat{q}_{i}p_{i}\)^{2n}\,,
\end{equation}
where $\hat{q}$ is a unit vector in the direction of $\vec{q}$. If equation \eqref{alt_In_def} is to decay with distance no faster than $r^{-1}$ when acting on the wave function $\psi$, given equation \eqref{psqr_posit}, the integral must contain a term proportional to $\vec{p}^{\,2n}=p_r^{\,2n}+\dots$. This requires the following contribution to \eqref{alt_In_def} to be nonzero
\begin{eqnarray}\label{script_I}
\mathbb{I}^{(n)} &=& \f{1}{4\pi^2 r} \int_0^\infty dw \int_{-1}^1 dx \,e^{i w x} \(\hat{q}_{r} p_r\)^{2n}\notag\\
&=&\f{1}{4\pi^2 r}\(p_r\)^{2n}\int_0^\infty dw \int_{-1}^1 dx\, \cos{\(w x\)}\, x^{2n}\,,
\end{eqnarray}
where the substitution $\hat{q}_{r}=\cos{\theta}=x$ has been made and the symmetry of the $x$ integral has been exploited. A series representation of the cosine may be used to show that
\begin{multline}
\int_{-1}^1 dx\, \cos{\(w x\)}\, x^{2n}=\\
\f{1}{n+1/2}{_1}F_2\[n+\frac{1}{2}; \frac{1}{2} , n+ \frac{3}{2}; -\frac{w^2}{4}\]\,,
\end{multline}
where ${_1}F_2$ is a hypergeometric function. With the variable substitution $\chi=\tfrac{w^2}{4}$, it follows that
\begin{eqnarray}
\mathbb{I}^{(n)} &=& \f{1}{4\pi^2 r}\(p_r\)^{2n} \int_0^\infty \, \f{d\chi\,\chi^{-1/2}}{n+1/2} {_1}F_2\[n+\frac{1}{2}; \frac{1}{2} , n+ \frac{3}{2}, -\chi\]\notag\\
&=& \f{1}{4\pi^2 r}\(p_r\)^{2n}\Gamma\(\frac{1}{2}\)\lim_{\mu\to0^+} \mu^n \gamma\(n+\frac{1}{2},\f{1}{\mu}\)\notag\\
&=& \f{1}{4\pi^2 r}\(p_r\)^{2n}\Gamma\(\frac{1}{2}\)\lim_{\mu\to0^+}  \mu^n \Gamma\(n+\f{1}{2}\)
\end{eqnarray}
where, in the second line, a known integral \cite{gradshteyn2014table} was used in conjunction with the series representation of the incomplete gamma function, $\gamma(\a,x)$, and in the third line use was made of the limiting behavior
\begin{equation}
 \lim_{\mu\to0^+}\gamma\(n+\frac{1}{2},\f{1}{\mu}\) \sim \Gamma\(n+\f{1}{2}\) + {\cal O}\(e^{-1/\mu}\mu^{1/2-n}\)\,.
\end{equation}
It follows that
\begin{eqnarray}
\mathbb{I}^{(n\neq0)}&=&0\,,
\end{eqnarray}
therefore, $I^{(n\neq0)}$ acting on $\psi$ decays with distance faster than $1/r$ and hence only the $n=0$ term contributes to the effective potential a term proportional to $1/r$. After Fourier transforming equation \eqref{U_eff_q-space_02}, we learn
\begin{equation}\label{U_eff}
\lim_{r\to\infty} U_\text{eff}(r) = -\f{Z\alpha}{r}                        \(1+\f{\vec{p}^{\,2}}{\vep_1\vep_2}   \)
\end{equation}
a result that would also follow from setting by hand $\omega=0$ in equation \eqref{U_gen_q-space}.



The effective commutation relation  \eqref{commutation_simplification} may be used to rearrange \eqref{SchrodingerEqn} algebraically to yield
\begin{equation}
\vec{p}^{\,\,2}\psi=\(-q^2 + \f{2\hat{m}Z\alpha}{r} +\f{{\cal O}\(\a^2\)}{r^2}\)\psi\,,
\end{equation}
which is superficially identical to the canonical Schrodinger-Coulomb equation. However, here 
\begin{equation}\label{qsqr_eqn}
q^2=-\frac{E\(E+2m_1\)\(E+2m_2\)\(E+2m_1+2m_2\)}{4\(E+m_1+m_2\)^2}
\end{equation}
and
\begin{equation}\label{mhat_eqn}
\hat{m}=\f{E^2+2m_1m_2+2E\(m_1+m_2\)}{2\(E+m_1+m_2\)}\,.
\end{equation}

The radial solutions to the Schrodinger-Coulomb equation involve the superposition of two independent (regular and irregular) confluent hypergeometric functions. Here we focus exclusively on bound states ($E<0$), for which the normalizeable\footnote{I continue to be grateful to Harsh Mathur for emphasizing the importance of seeking solutions that are normalizeable in the $r\to \infty$ limit.} radial solutions are
\begin{equation}\label{radial_Solution}
R(r) = e^{-qr} r^{\tilde{\ell} } \, U\(1+\tilde{\ell} -\f{\hat{m}Z\alpha}{q} , 2(\tilde{\ell}+1) , 2qr\)  \,,
\end{equation}
where $U(a,c,x)$ is Tricomi's confluent hypergeometric function,
\begin{equation}
\tilde{\ell} = \ell + \frac{{\cal O}\(\a^2\)}{2\ell+1}
\end{equation}
and
\begin{equation}\label{q_eqn}
q=\f{\hat{m}Z\alpha}{n_\star}\,,
\end{equation}
where $n_\star$ is the effective quantum number. Two remarks are here warranted. Firstly, because the quantity $1+\tilde{\ell} -\f{\hat{m}Z\alpha}{q}$ will never be exactly equal to a negative integer in a real system, the regular radial solution diverges exponentially at large $r$ and cannot be normalized, therefore it has been discarded.  Secondly, the singular behavior of the solution involving the (irregular) Tricomi function near $r=0$ is not a concern with this approach because we would not, for example, try to normalize wavefunctions on the entire domain $0\leq r<\infty$. A Hartree-like approach \cite{hartree_1928} may be employed in which these exterior solutions are matched at a finite radius onto appropriate ``interior" wavefunctions. Alternatively, we can treat this radius as a system boundary and posit that boundary conditions encode information about the omitted interactions in the vicinity of the origin \cite{Jacobs:2019woc}.

The solutions here are nevertheless thought to be in some sense near to their ``pure relativistic Coulomb" forms, which are regular at $r=0$, and which corresponds to $n_\star$ being is equal to a positive integer, $n$. As is standard, we write
\begin{equation}
n_\star=n-\de\,,
\end{equation}
where the quantum defect, $\de$, accounts for the shorter-ranged interactions that have not been modeled explicitly \cite{hartree_1928,Seaton_1983,Jacobs:2019woc}. From equations \eqref{qsqr_eqn}, \eqref{mhat_eqn}, and \eqref{q_eqn} it follows that the physical energy eigenvalues are
\begin{equation}\label{general_E_sol}
E=\sqrt{m_1^2+m_2^2+\f{2m_1m_2}{\sqrt{1+\(\f{Z\alpha}{n_\star}\)^2}}} - \Big(m_1+m_2\Big)\,.
\end{equation}
Equation \eqref{general_E_sol} can also be obtained as a special case\footnote{I am grateful to an anonymous referee for pointing out this existing body of work on the relativistic two-body problem.} of the analysis in Ref \cite{Crater:1992xp}, a point that is addressed in Appendix \ref{Crater_et_al}. However, it appears not to have been appreciated in Ref \cite{Crater:1992xp} that this result is exact for the electromagnetic interaction in the large $n$ (large $r$) limit, and hence may be used as the starting point of a relativistic effective atomic theory.

To see that the energies in  \eqref{general_E_sol} are plausible, a Taylor expansion in small $Z\alpha/n_\star$ yields a series whose first term is the Rydberg-Ritz formula, equation \eqref{NR_Ritz}, followed by  relativistic kinetic correction terms,
\begin{widetext}
\begin{equation}\label{general_E_sol_expanded}
E=-\f{1}{2}\f{m_1 m_2}{m_1+m_2} \(\f{Z\a}{n_\star}\)^2
\(1 - \f{3m_1^2+5m_1m_2+3m_2^2}{4\(m_1+m_2\)^2}\(\f{Z\a}{n_\star}\)^2  + {\cal O}\(\f{Z\a}{n_\star}\)^4\) \,. 
\end{equation}
\end{widetext}
~\\
In the limit $m_1/m_2\to 0$, equation \eqref{general_E_sol} becomes the canonical Dirac-Coulomb energy levels of equation \eqref{canonical_Dirac-Coulomb}. In the case of $m_1=m_2=m$, equation \eqref{general_E_sol} is consistent with the results of \cite{Jacobs:2021xgp}, namely
\begin{equation}
E=m\(\sqrt{2+\f{2}{\sqrt{1+\(\f{Z\alpha}{n_\star}\)^2}}} - 2\)\,,
\end{equation}

\ssec{Relativistic Ritz-like Defect Theory}
Finally, we must consider the form of the quantum defect. In a previous article concerning the special case of positronium \cite{Jacobs:2021xgp}, a Ritz-like expansion was posited, namely a series expansion in which the energies are assumed to be small relative to some high energy scale, $\Lambda$:
\begin{equation}\label{defect_ansatz}
\de_{\ell s j}=\de_{(0)\ell s j}+\l_{(1)\ell s j} \f{E}{\L}+\l_{(2)\ell s j}\(\f{E}{\L}\)^2+\dots\,,
\end{equation}
where, for a non-annihilating system, the defect parameters, $\de_{(0)\ell s j}, \l_{(1)\ell s j}, \l_{(2)\ell s j},\,\dots$ are real and should depend on the orbital ($\ell$), total spin ($s$), and total ($j$) angular quantum numbers of the state; alternatively, the system can be described by the orbital, total ``electronic", and total system quantum numbers, $\ell$, $j$, and $f$, respectively. A \emph{modified} ansatz  written as a series in inverse powers of $(n-\de_{(0)})$ is equivalent and is significantly easier to use for data fitting. Analyzing the asymptotic (large $n$) behavior of \eqref{general_E_sol}, it may be verified that
\begin{widetext}
\begin{multline}\label{defect_modified_ansatz}
\de_{\ell s j}=\de_{(0)\ell s j}+ \f{\de_{(2)\ell s j}}{\(n-\de_{(0)\ell s j}\)^2}+\f{\de_{(4)\ell s j}}{\(n-\de_{(0)\ell s j}\)^4} +\f{2\de_{(2)\ell s j}^2}{\(n-\de_{(0)\ell s j}\)^5}  +\f{\de_{(6)\ell s j}}{\(n-\de_{(0)\ell s j}\)^6}  +\f{6\de_{(2)\ell s j} \de_{(4)\ell s j}}{\(n-\de_{(0)\ell s j}\)^7}  +\f{\de_{(8)\ell s j}}{\(n-\de_{(0)\ell s j}\)^8} \\
+\f{4\de_{(4)\ell s j}^2 + 8\de_{(2)\ell s j} \de_{(6)\ell s j}}{\(n-\de_{(0)\ell s j}\)^9}
 +\f{\de_{(10)\ell s j}}{\(n-\de_{(0)\ell s j}\)^{10}} +\f{-40\de_{(2)\ell s j}^4 + 10\de_{(4)\ell s j} \de_{(6)\ell s j}+ 10\de_{(2)\ell s j} \de_{(8)\ell s j}}{\(n-\de_{(0)\ell s j}\)^{11}}+\dots\,.
\end{multline}
\end{widetext}
This form differs from that hitherto used widely in the literature, equation \eqref{modified_NR_Ritz}, because \eqref{defect_modified_ansatz} contains all inverse powers of $(n-\de_{(0)})$ beyond 4th order, a result that follows from the low-energy expansion of the defect. In fact, it may be verified that \eqref{defect_modified_ansatz} applies even when using the nonrelativistic energy formula, equation \eqref{NR_Ritz}.

 Equation \eqref{general_E_sol} and the modified defect ansatz in \eqref{defect_modified_ansatz} are foundational for the remainder of this article. Given that it has a proper relativistic foundation, the use of these formula will here be referred to as the \emph{relativstic Ritz} approach.

\sec{Fits to theoretical energy levels of hydrogen}\label{QED_level_match}

Analyzing the bound-state energy levels predicted by BSQED with formulas  \eqref{NR_Ritz} and  \eqref{general_E_sol} is perhaps the most expedient way to establish superiority of the latter relativistic formula. A tabulation of bound-state energies of hydrogen ($Z=1)$ by Horbatsch and Hessels \cite{horbatsch2016tabulation}, which includes hyperfine structure, will be used as a fiducial set of data for this purpose. I will analyze here a selection of the $nS$, $nP$, and $nD$ energy levels of hydrogen, which may be found in Tables III, IV, and V of Ref. \cite{horbatsch2016tabulation}.


Here we reintroduce factors of Planck's constant and the speed of light. Prior to fitting it is important to note that in Ref. \cite{horbatsch2016tabulation} the following parameter values were used:
\begin{eqnarray}
m_p/m_e&=&1836.15267389(17)\\
\a^{-1}&=&137.035999139(31)\label{HorbHessAlphainv}\\
\f{m_ec^2 \a^2}{2h}\equiv cR_\infty&=&3\,289\,841\,960\,248.9(3.0)\,\text{kHz}\,.
\end{eqnarray}
It will therefore be consistent to use as input values for the electron and proton masses
\begin{equation}
\f{m_e c^2}{h}=1.235\,589\,964\,81\times10^{17}\,\text{kHz}
\end{equation}
and
\begin{equation}
\f{m_p c^2}{h}=2.268\,731\,817\,71\times10^{20}\,\text{kHz}\,.
\end{equation}

Any fit requires truncation of equation \eqref{defect_modified_ansatz}. At lowest order (LO), only $\a$ and the defect parameter $\de_{(0)\ell j f}$ is used in a numerical fit; at next-to-leading order (NLO) $\de_{(0)\ell j f}$ and $\de_{(2)\ell j f}$ are used, and so on. This means that at order N$^b$LO there are $b+2$ fit parameters. Fitting was computed using the Levenburg-Marquardt algorithm, weighting each data point interval by the inverse-square of its uncertainty. As a practical matter regarding numerical precision, it was necessary to first expand equation \eqref{general_E_sol} in powers of $\a$ before applying the  N$^5$LO fit; an expansion to order $\a^8$ was deemed sufficient as it showed negligible changes to the fit parameters and various goodness-of-fit measures compared to the order $\a^6$ expansion. The same was found to be true for all other lower order fits.

\ssec{Comparing the nonrelativistic to the relativistic Ritz approach}

\begin{figure*}[h!]
\begin{tabular}{cc}
  \includegraphics[width=85mm]{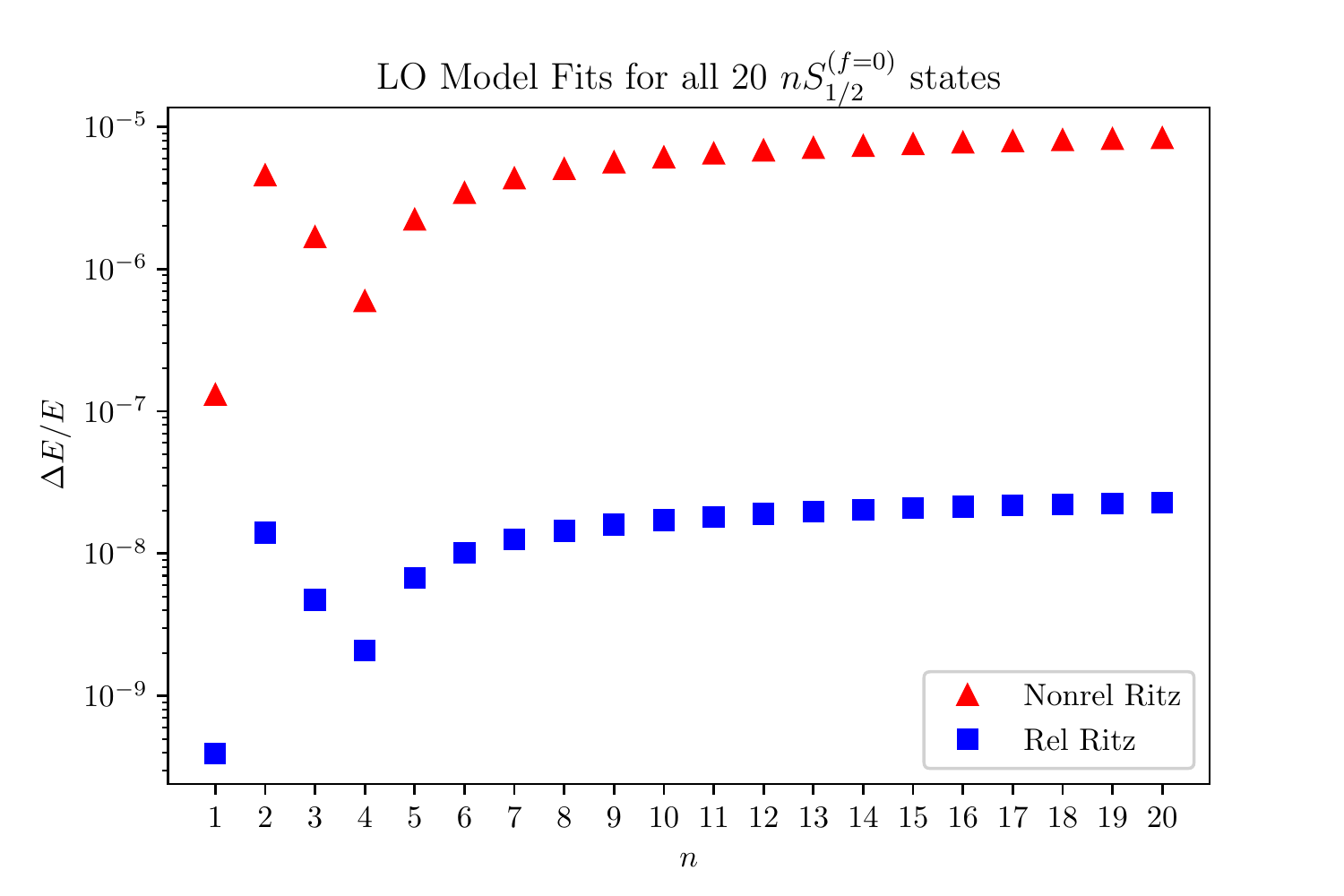} &   \includegraphics[width=85mm]{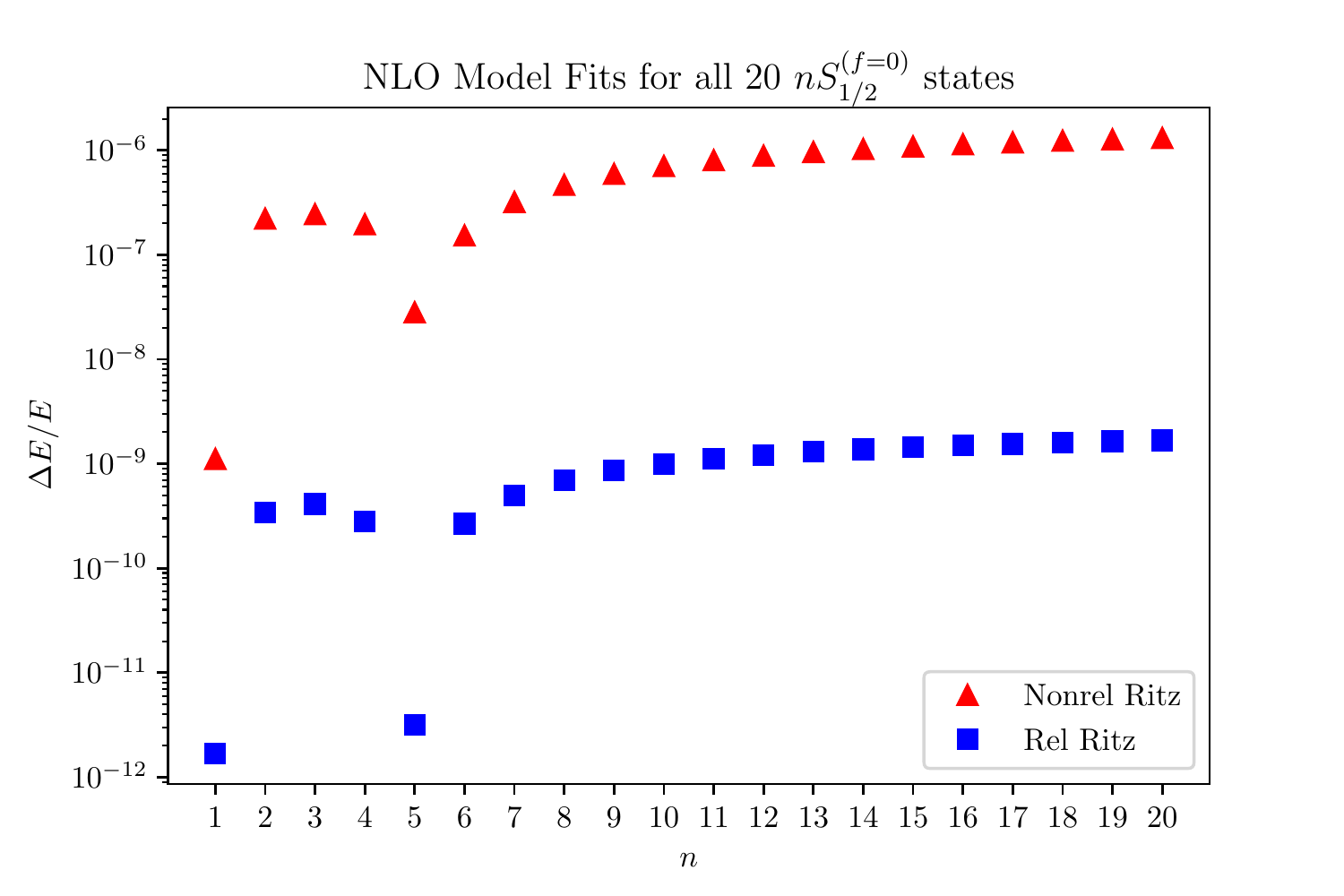} \\
(a) LO & (b) NLO \\[6pt]
 \includegraphics[width=85mm]{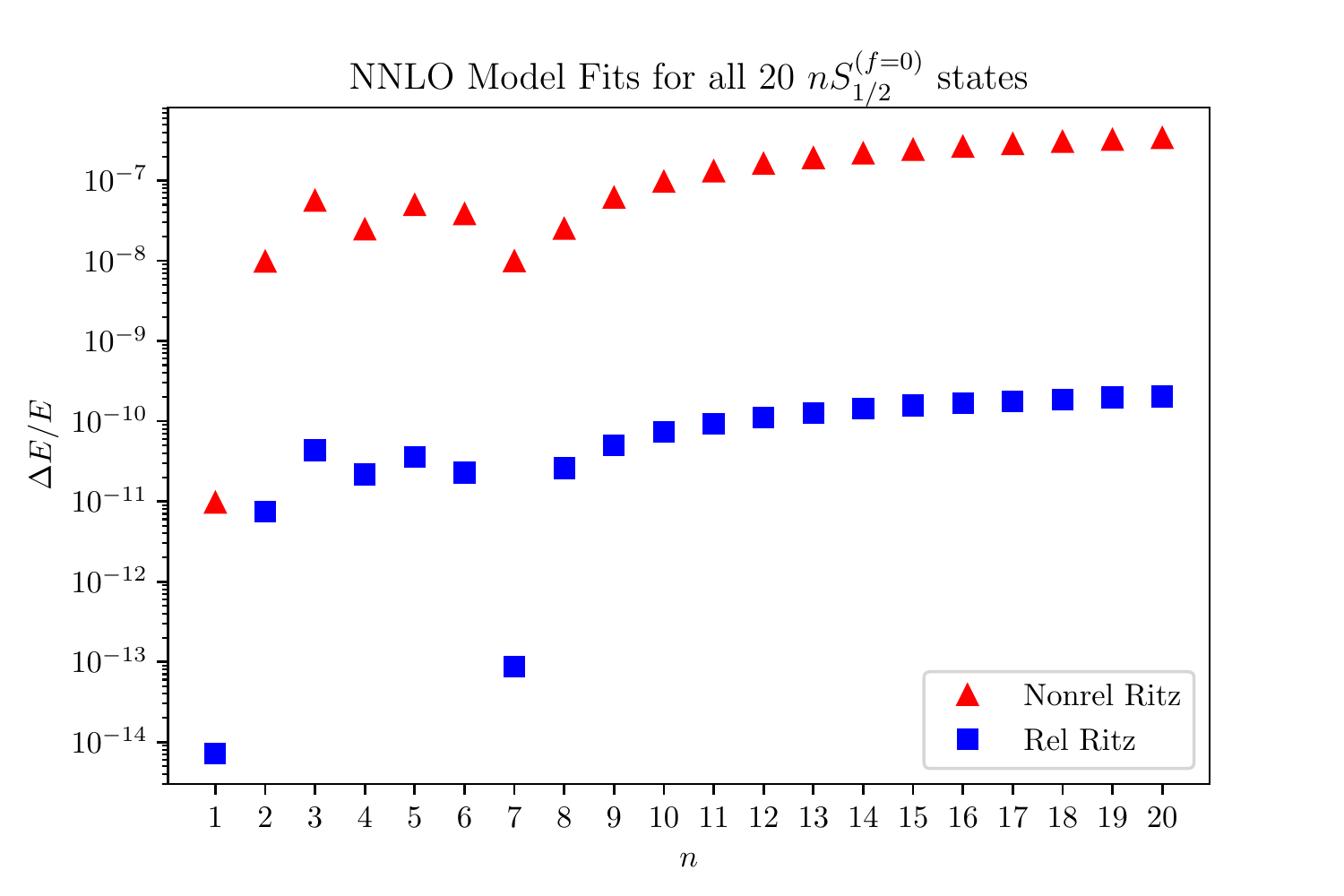} &   \includegraphics[width=85mm]{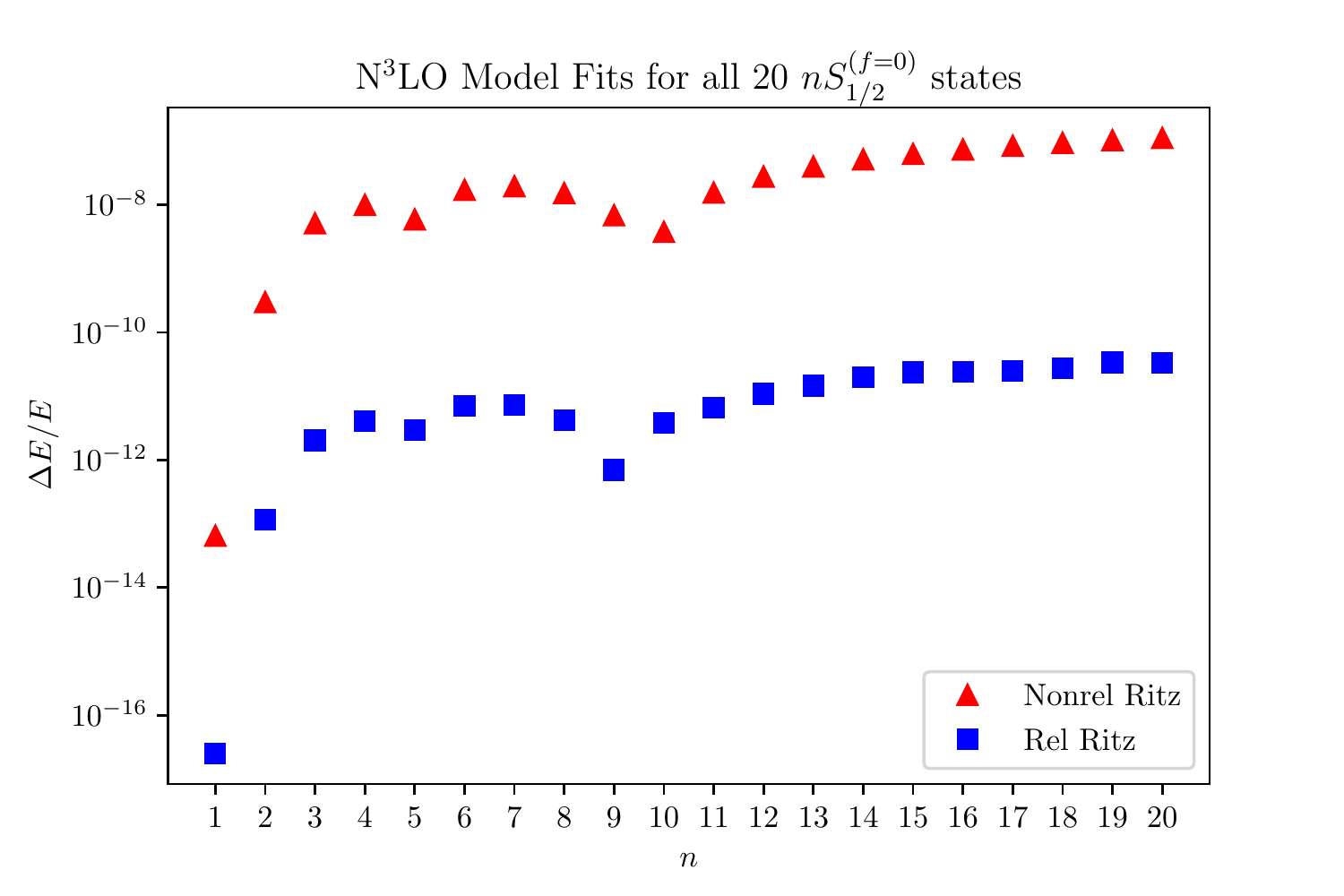} \\
(c) NNLO & (d) N$^3$LO \\[6pt]
 \includegraphics[width=85mm]{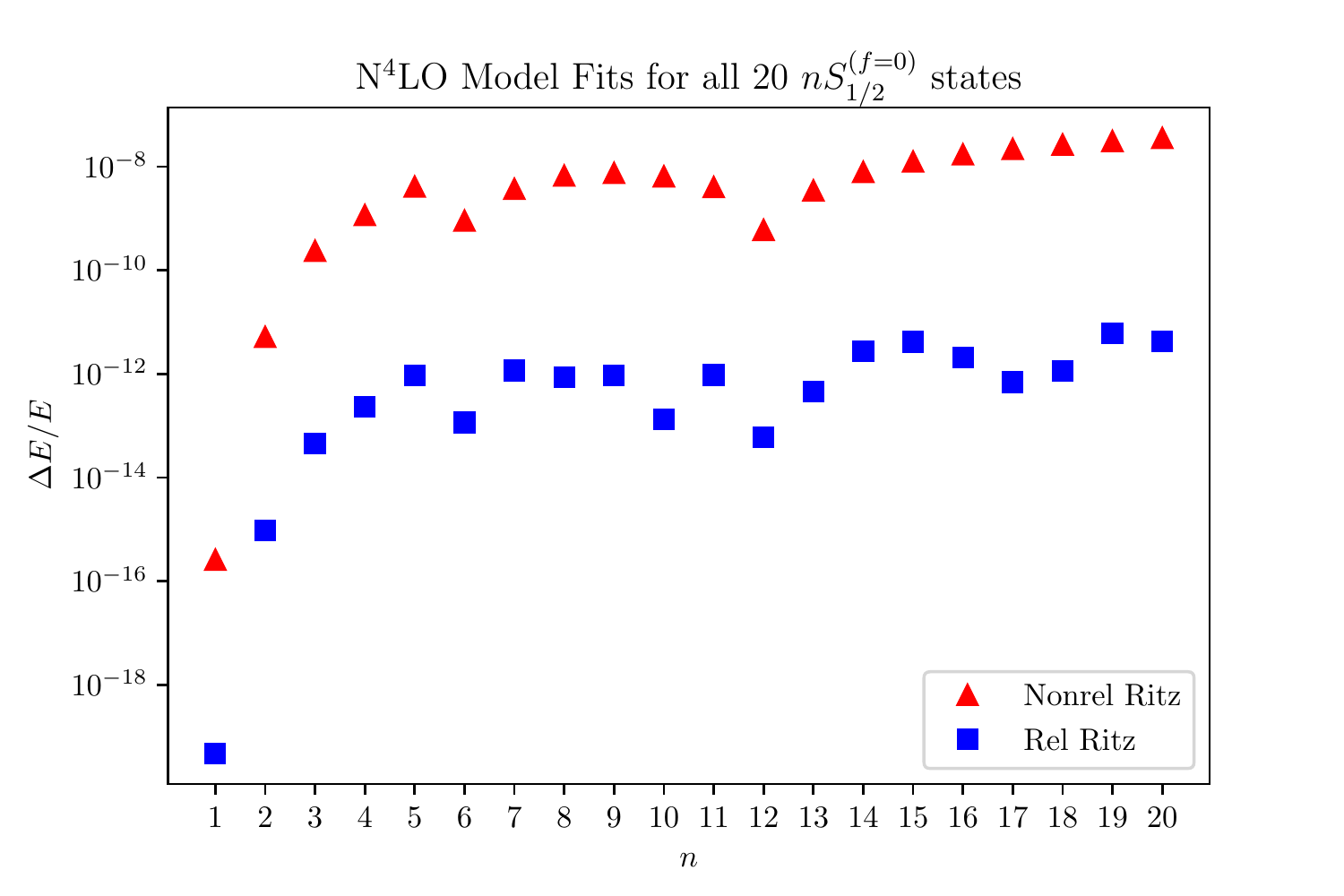} &   \includegraphics[width=85mm]{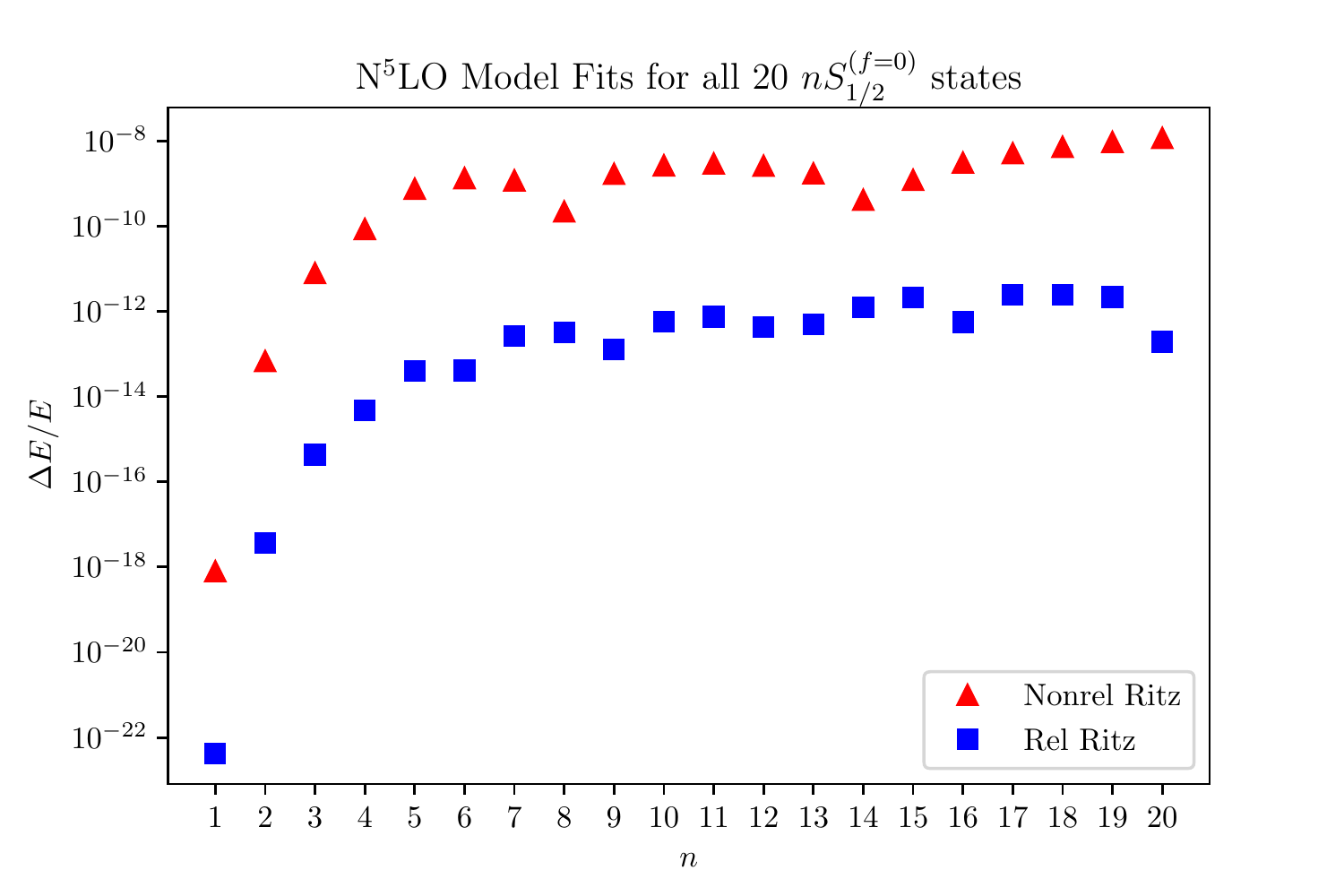} \\
(e) N$^4$LO & (f) N$^5$LO \\[6pt]
\end{tabular}
\caption{\raggedright Relative residuals from the fits to all 20 of the $nS_{1/2}^{(f=0)}$ QED-predicted energy levels reported in Ref \cite{horbatsch2016tabulation}. Each graph corresponds to a different perturbative order. The relative residuals from the nonrelativistic Ritz formula \eqref{NR_Ritz}  are indicated with red triangles and those from the relativistic Ritz formula \eqref{general_E_sol}  are in blue squares. }
\label{Rel_res_all_S_20}
\end{figure*}

The results from fitting equations \eqref{NR_Ritz}  and \eqref{general_E_sol}  to  all 20 of the $nS_{1/2}^{(f=0)}$ states is summarized below; fit values for $\a^{-1}$ are displayed in Table \ref{Fit_alphainv_Sf0_all_20} and the relative residuals from each fitting formula are displayed in Figure \ref{Rel_res_all_S_20}. It should be noted that the $nS_{1/2}^{(f=0)}$ energies reported in \cite{horbatsch2016tabulation} are given at a precision level no better than one part per $10^{12}$ for $n\geq6$.

\begin{table}[H]
\centering
\caption{Values of $\a^{-1}$  from model fits to BSQED-predicted energy levels of the $nS_{1/2}^{f=0}$ states of atomic hydrogen. Values should be compared with the input value of $\a^{-1}=137.035\,999\,139(31)$ used in Ref. \cite{horbatsch2016tabulation}.}
\begin{tabular}{c l l}  \hline\hline
Order &  NR Ritz $\a^{-1}$&  Rel. Ritz $\a^{-1}$   \\ \hline
LO &$137.035\,27(7)$ & $137.036\,001\,1(2)$ \\
NLO &$137.035\,86(1)$ & $137.035\,999\,31(2)$ \\
NNLO &$137.035\,949(5)$ & $137.035\,999\,165(3)$  \\
N$^3$LO &$137.035\,975(2)$ & $137.035\,999\,1449(8)$ \\
N$^4$LO &$137.035\,9859(9)$ & $137.035\,999\,1404(1)$  \\
N$^5$LO &$137.035\,9911(5)$ & $137.035\,999\,1395(1)$  \\  \hline\hline
\end{tabular}
\label{Fit_alphainv_Sf0_all_20}
\end{table}


\ssec{Fitting and predicting with subsets of levels}

To compare fit results amongst different angular momentum channels on equal footing, the first 7 levels of the $nS_{1/2}^{(f=0)}$, $nS_{1/2}^{(f=1)}$, $nP_{1/2}^{(f=0)}$, $nP_{1/2}^{(f=1)}$, $nD_{3/2}^{(f=1)}$ and $nD_{3/2}^{(f=2)}$ states are fit with equation \eqref{general_E_sol} at various orders in the defect expansion, and extrapolations are made to the remaining levels reported in the corresponding tables in Ref. \cite{horbatsch2016tabulation}. The notation used for the defect parameters is $\de_{(i)\ell j f}$.

\begin{figure*}[hp]
\begin{tabular}{cc}
  \includegraphics[width=86mm]{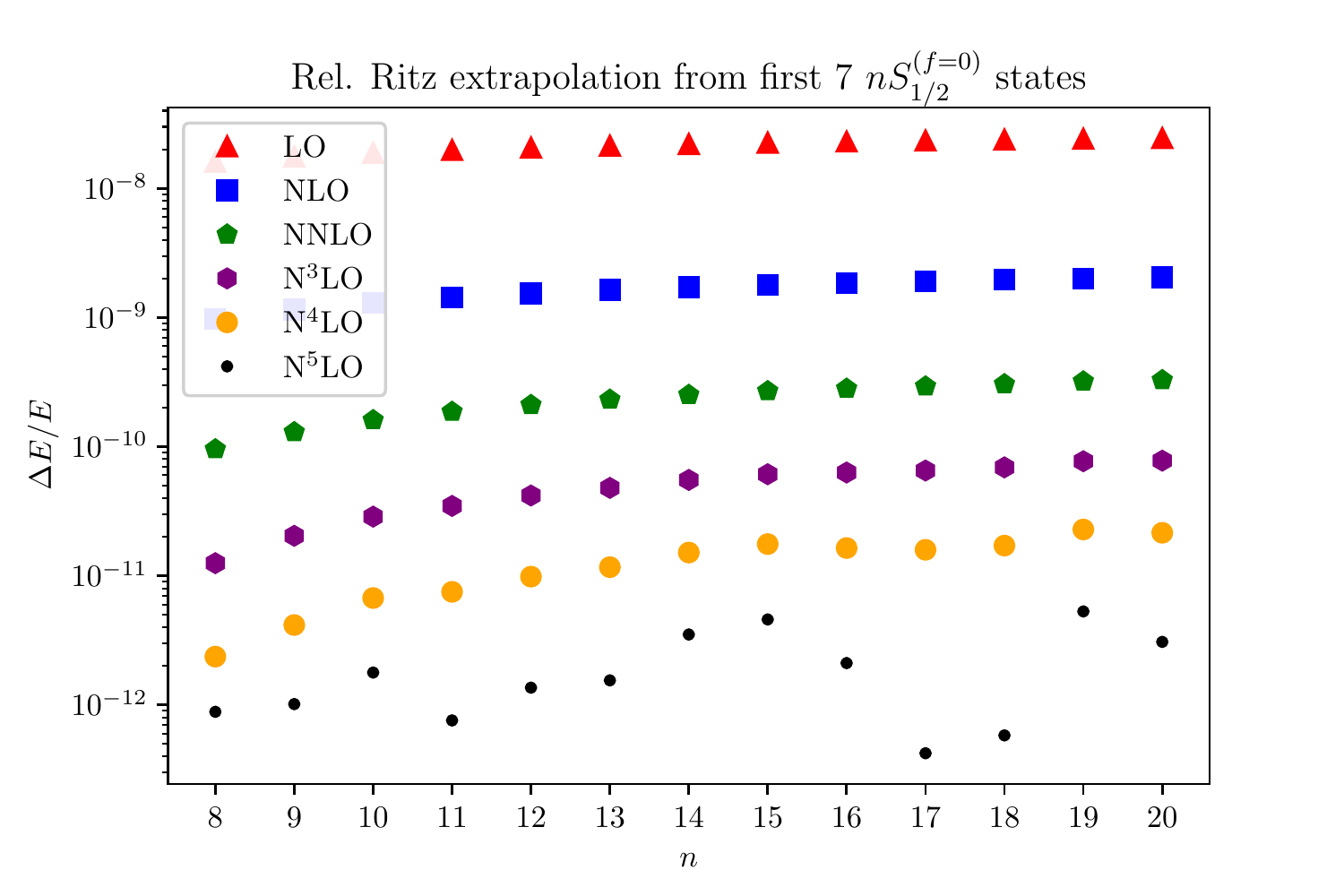} &   \includegraphics[width=86mm]{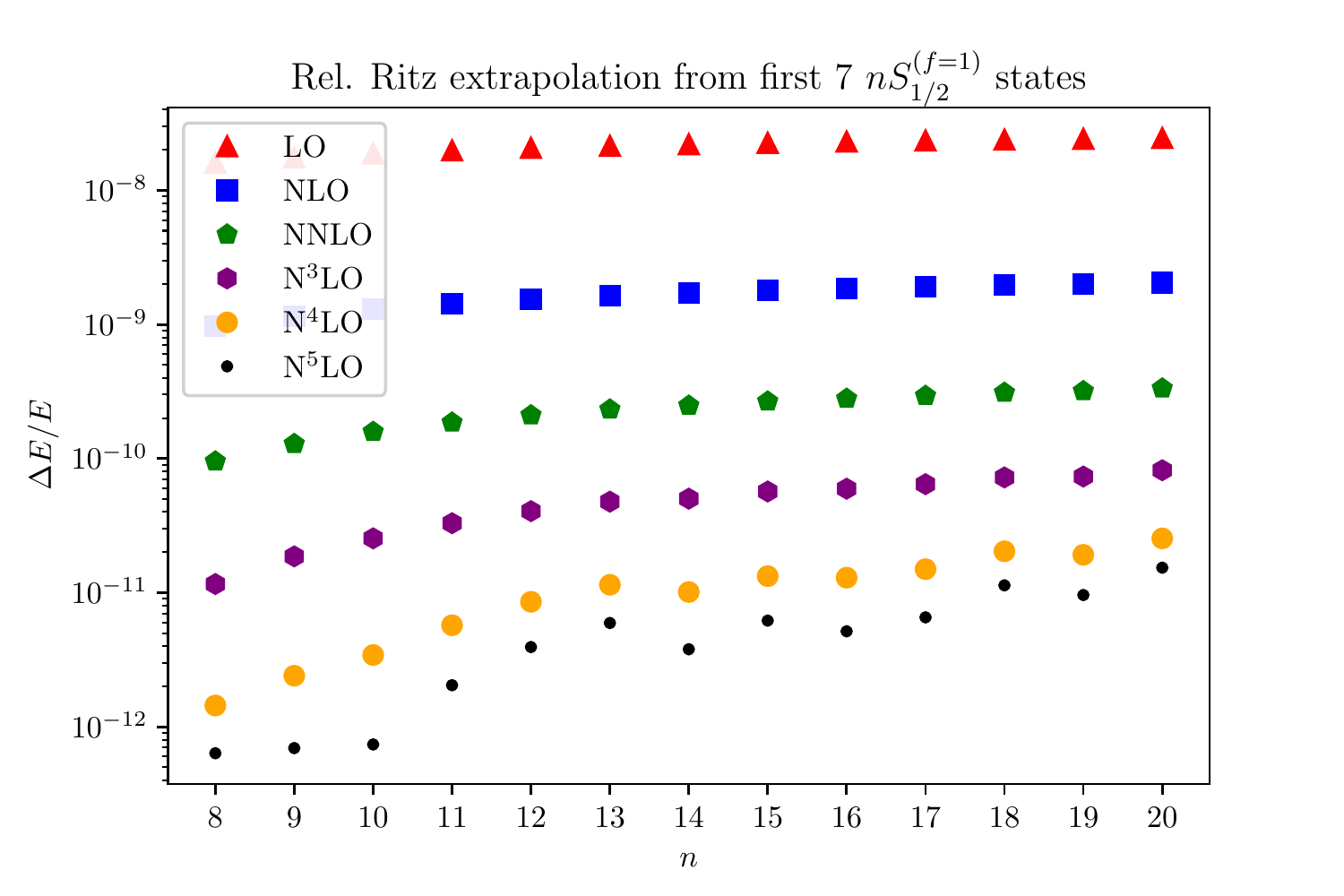} \\
(a) $nS_{1/2}^{(f=0)}$ & (b) $nS_{1/2}^{(f=1)}$ \\[6pt]
 \includegraphics[width=86mm]{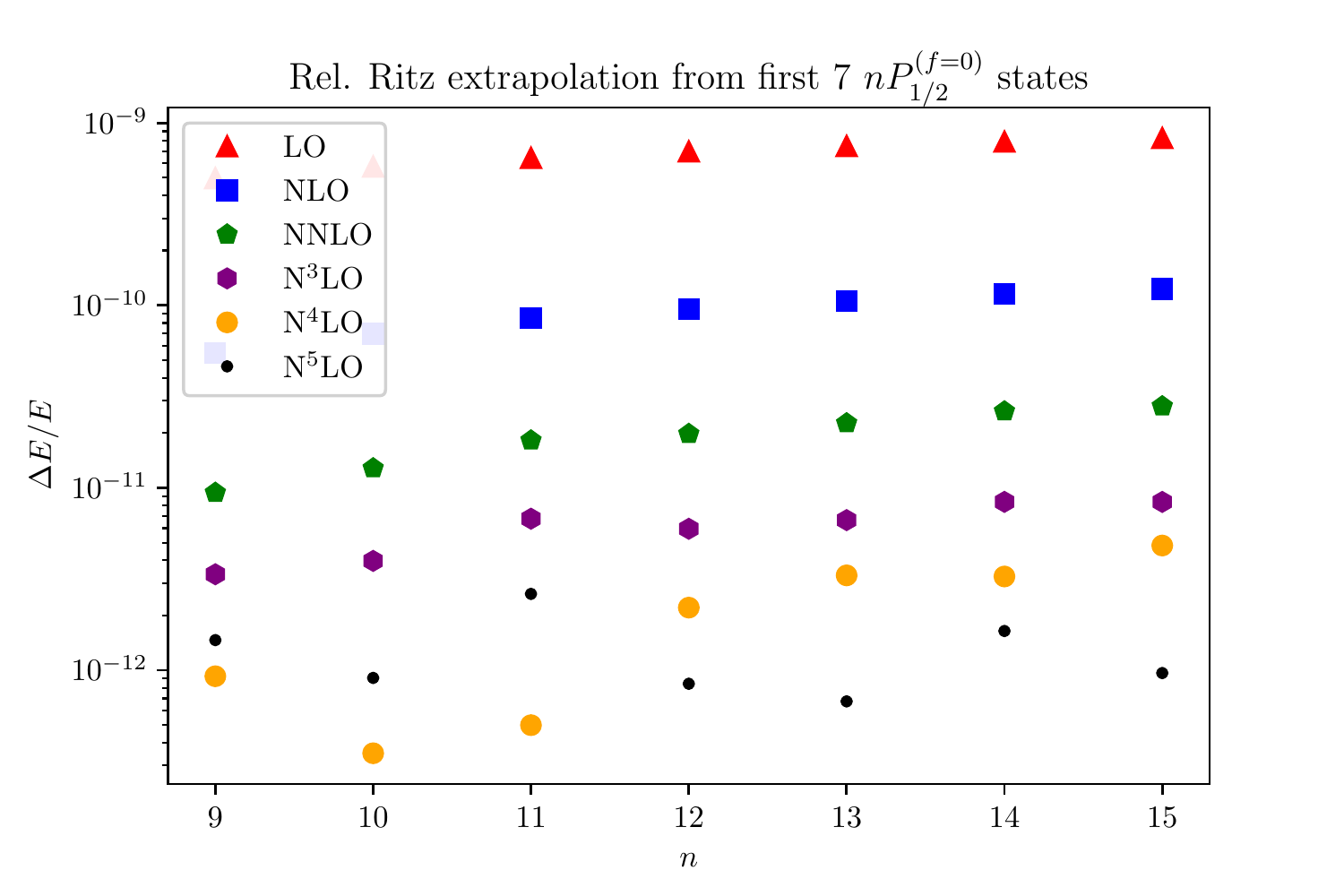} &   \includegraphics[width=86mm]{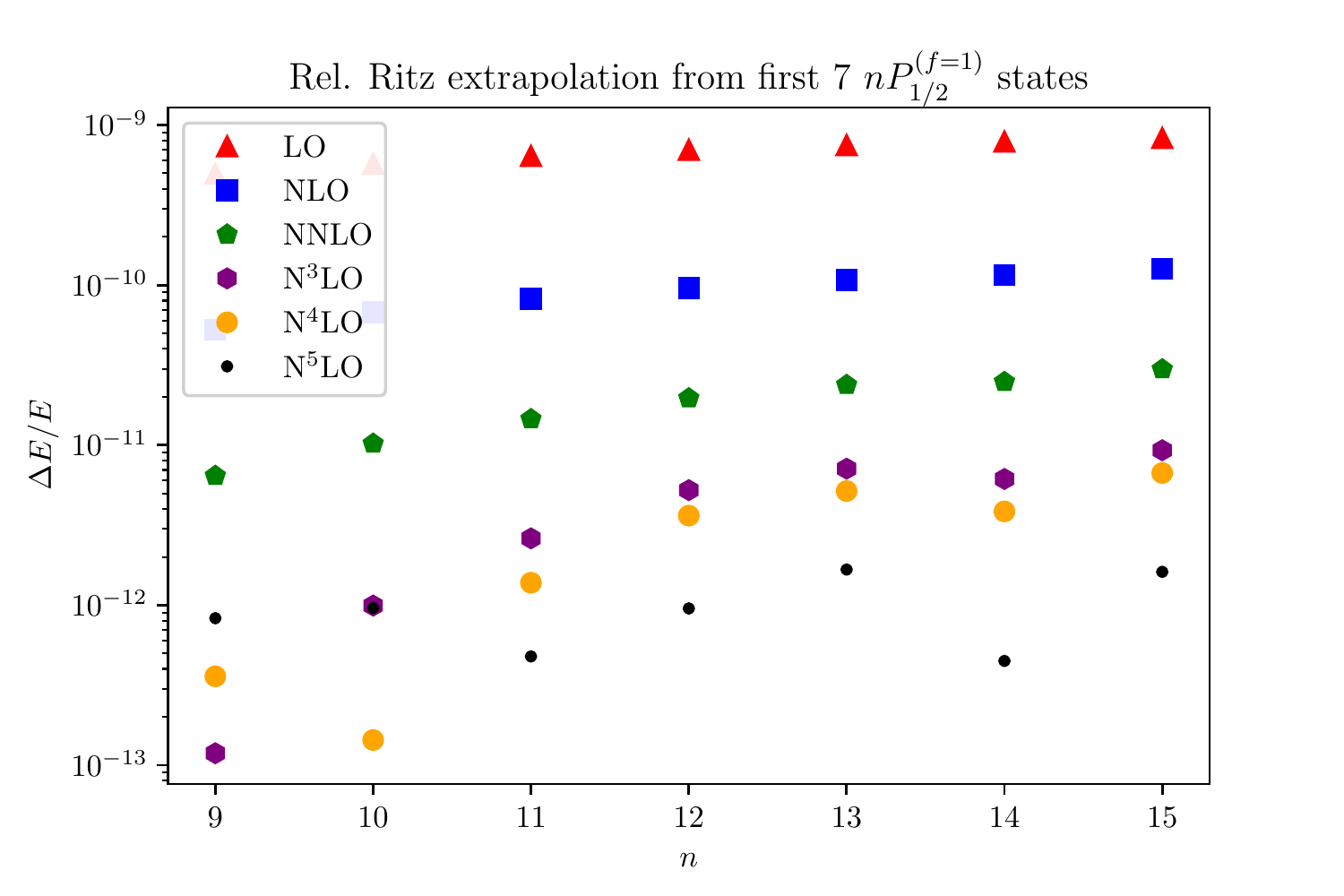} \\
(c) $nP_{1/2}^{(f=0)}$ & (d) $nP_{1/2}^{(f=1)}$ \\[6pt]
 \includegraphics[width=86mm]{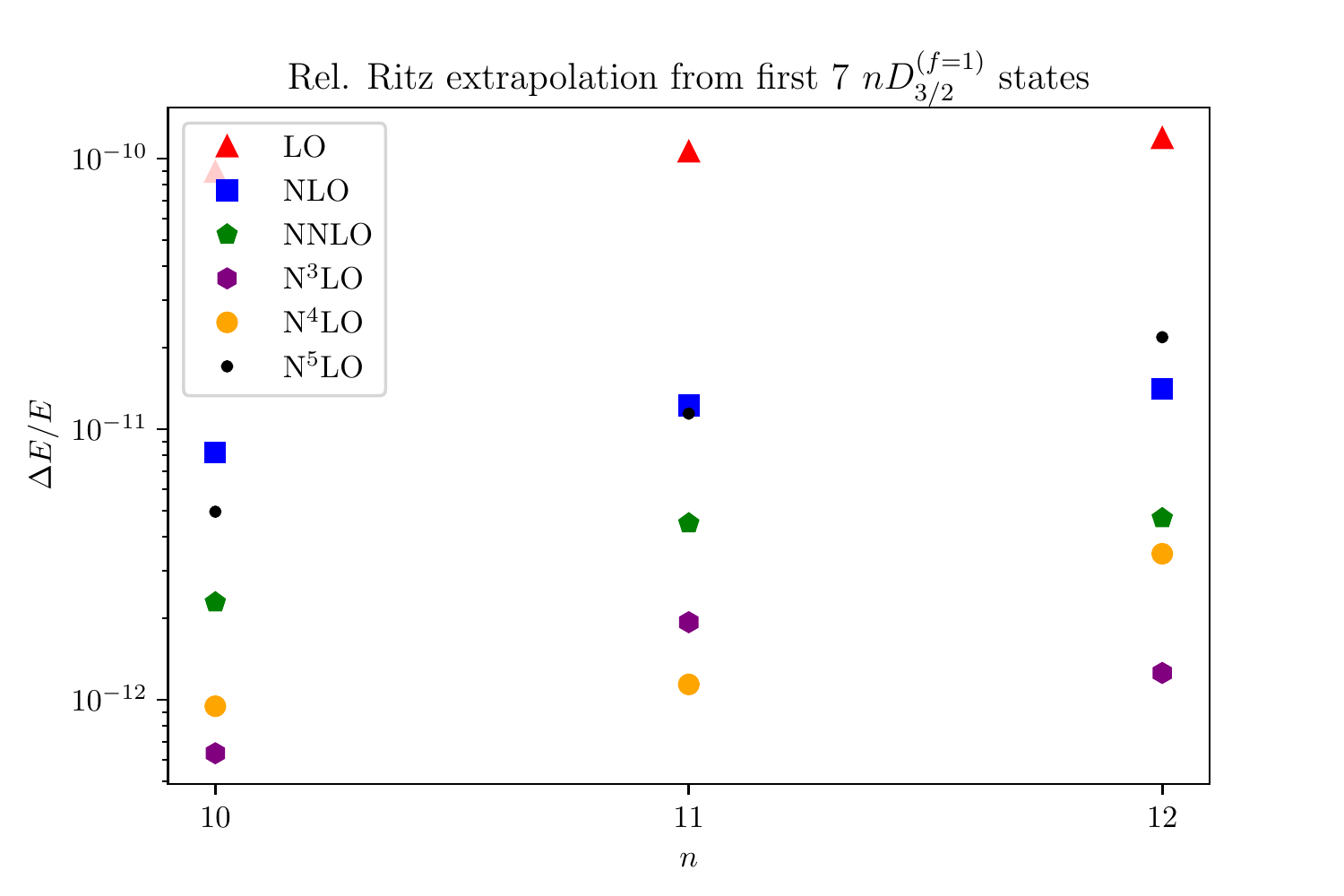} &   \includegraphics[width=86mm]{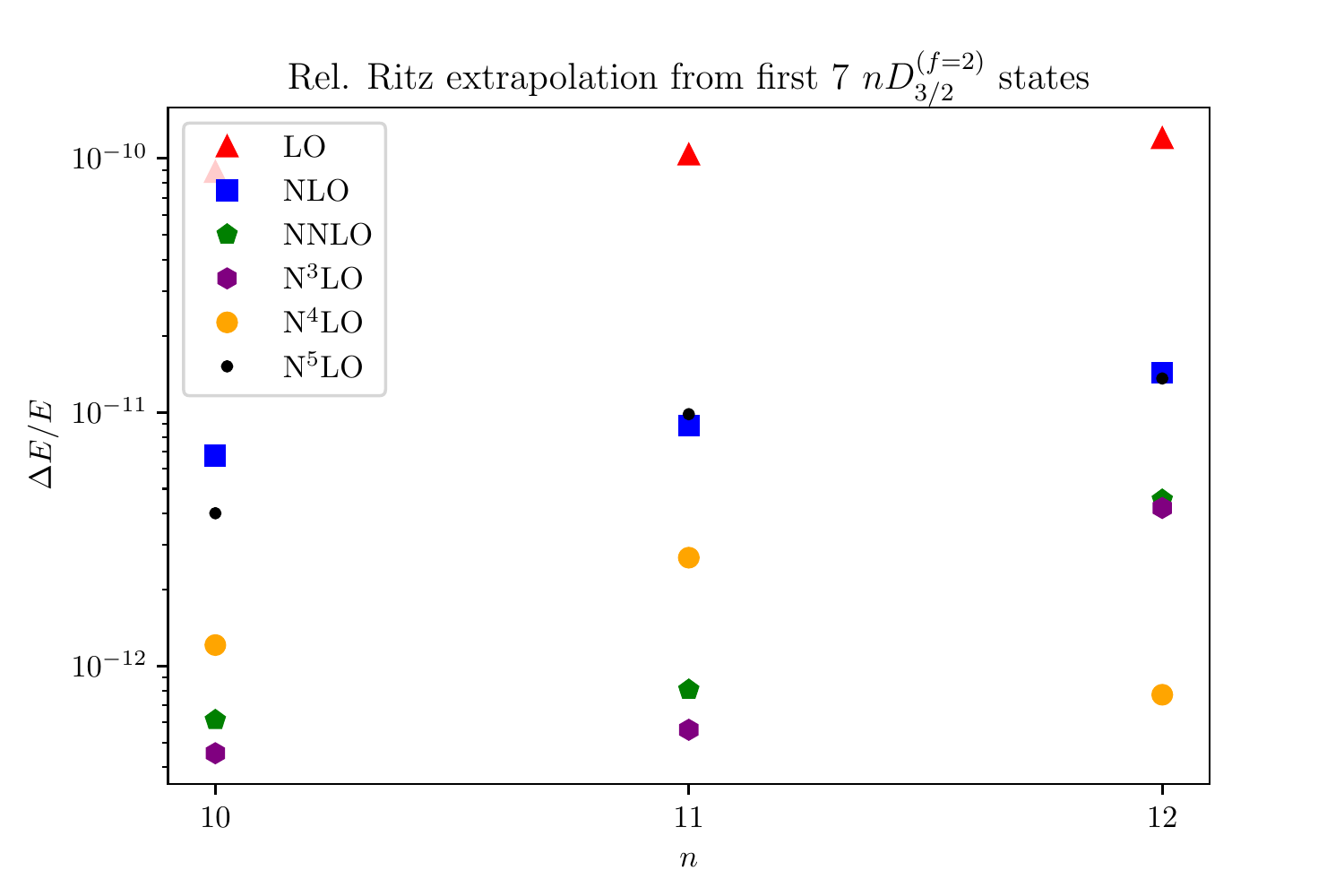} \\
(e) $nD_{3/2}^{(f=1)}$ & (f) $nD_{3/2}^{(f=2)}$ \\[6pt]
\end{tabular}
\caption{Relative residuals for predictions made from the fit first 7 states indicated.}
\label{Predictions_from_first_7_SPD}
\end{figure*}

Parameters values for a representative sample of the NNLO fits is shown in Tables \ref{table:fit_parameters_S_f0}, \ref{table:fit_parameters_P_f0}, and \ref{table:fit_parameters_D_f1}. The results in the form of relative residuals are also summarized in Figure \ref{Predictions_from_first_7_SPD}.  It is apparent that, with the exception of the $nS$ states, the inclusion of more parameters does not necessarily yield a better fit, i.e., a longer series expansion of the quantum defect parameter does not necessarily improve accuracy.  This may be the result of the fact that many of the values used for fitting were only reported to a relative precision no better than $10^{-12}$, so an improved fit is impossible.   On the other hand, the $nP$ residuals in Figure \ref{Predictions_from_first_7_SPD} suggest that at larger $n$ more parameters may be more appropriate. This is consistent with the idea that the quantum defect series should be treated an \emph{asymptotic series expansion} in small $E$, or large $n$, which does not necessarily converge. This motivates the analysis in the following section. Lastly, it is noteworthy that relative residuals in Figure \ref{Predictions_from_first_7_SPD} and the progressively accurate fit values for  $\a^{-1}$ in Tables \ref{table:fit_parameters_S_f0}, \ref{table:fit_parameters_P_f0}, and \ref{table:fit_parameters_D_f1} suggests that fewer defect parameters are needed for progressively higher orbital angular momentum states. This is consistent with the fact that those states are progressively less sensitive to interactions close to the origin.

In any case, not only does equation \eqref{general_E_sol} clearly provide a superior fit to the canonical Rydberg-Ritz formula \eqref{NR_Ritz},  it has the potential to provide a novel means of determining at least one physical parameter ($\a$) by applying it to actual transition data. It may also allow for consistency checks of data sets, which is of particular interest in light of possible correlated errors within the hydrogen and deuterium \cite{Khabarova:2021trt} data sets.

\begin{table}[h]
\centering
\caption{\raggedright Fit parameters from NNLO fit to the first 7 $nS_{1/2}^{(f=0)}$ levels from \cite{horbatsch2016tabulation} .}
\begin{tabular}{c | c}  \hline\hline
Parameter&  \\ \hline
$\a^{-1}$  & $137.035\,999\,176(7)$\\
$\de_{(0) 0\f{1}{2}0}$ & $2.55042(2)\times10^{-5}$\\  
$\de_{(2) 0\f{1}{2}0}$ & $5.61(7)\times10^{-8}$\\  
$\de_{(4) 0\f{1}{2}0}$ & $-1.48(6)\times10^{-8}$\\   \hline\hline
\end{tabular}
\label{table:fit_parameters_S_f0}
\end{table}

\begin{table}[h]
\centering
\caption{Fit parameters from NNLO fit, as in Table \ref{table:fit_parameters_S_f0}, for the $nP_{1/2}^{(f=0)}$ levels.}
\begin{tabular}{c | c}  \hline\hline
Parameter&  \\ \hline
$\a^{-1}$  & $137.035\,999\,1452(8)$\\
$\de_{(0) 1\f{1}{2}0}$ & $2.669299(4)\times10^{-5}$\\  
$\de_{(2) 1\f{1}{2}0}$ & $1.44(3)\times10^{-8}$\\  
$\de_{(4) 1\f{1}{2}0}$ & $-1.37(8)\times10^{-8}$\\   \hline\hline
\end{tabular}
\label{table:fit_parameters_P_f0}
\end{table}

\begin{table}[h]
\centering
\caption{Fit parameters from NNLO fit, as in Table \ref{table:fit_parameters_S_f0}, for the $nD_{3/2}^{(f=1)}$ levels.}
\begin{tabular}{c | c}  \hline\hline
Parameter&  \\ \hline
$\a^{-1}$  & $137.035\,999\,1410(3)$\\
$\de_{(0) 2\f{3}{2}1}$ & $1.332842(2)\times10^{-5}$\\  
$\de_{(2) 2\f{3}{2}1}$ & $6.9(3)\times10^{-9}$\\  
$\de_{(4) 2\f{3}{2}1}$ & $-1.2(2)\times10^{-8}$\\   \hline\hline
\end{tabular}
\label{table:fit_parameters_D_f1}
\end{table}




\sec{Insights into Bound-State QED}\label{QED_insights}

The remarkable success of the relativistic Ritz approach to characterize the BSQED energy levels, as described in the previous section, warrants additional investigation. To this end, we here consider exclusively non-annihilating systems, hence the energies are taken to be real\footnote{The case of positronium, an unstable system known to decay by annihilation,  was considered in \cite{Jacobs:2021xgp}.}. Let $m_1$ and $m_2$ be the mass of the lighter and heavier particles, respectively, and the dimensionless quantity
\begin{equation}
\mu = \(1+\f{m_1}{m_2}\)^{-1}\,,
\end{equation}
which takes a value between $1/2$ and 1, is used to quantify the recoil effect. The BSQED perturbative expression for the energy levels can be found in various publications, e.g. \cite{horbatsch2016tabulation, IEIDES200163, mohr2016codata}. Cumbersome as they may appear, their form simplifies asymptotically with increasing $n$. The mathematical relationship
\begin{equation}
\sum_{i=1}^n \frac{1}{i} = \psi(n) +\f{1}{n} + \gamma \,,
\end{equation}
where $\g$ is the Euler-Mascheroni constant and the digamma function, $\psi(n)\equiv \f{d}{dx}\ln{\Gamma{(x)}}$, is useful in what follows, especially when also noting the asymptotic behavior
\begin{equation}
\psi(n) \sim \ln{n}-\f{1}{2n} -\f{1}{12n^2}+{\cal O}\(n^{-4}\)\,.
\end{equation}
The so-called Bethe logarithm that appears in the energy expression, which will be discussed more in the following section, displays an asymptotic behavior that may be fit with a series expansion in inverse powers of $n$, as well; see, e.g., \cite{PhysRevA.72.012110}). It follows that the energy levels may be written in following the asymptotic (large $n$) form,
\begin{equation}\label{E_QED_asymp_form}
E\sim \sum _{A,N=2}^\infty C_{A,N}  \f{\m\,m_1\(Z\a\)^A}{n^N} \,,
\end{equation}
where the $C_{A,N}$ may contain logarithms of $\a$ and/or powers of $Z$. Using the known BSQED results
\begin{align}
C_{A,2}&=0 ~~~ (A>2)\\
C_{2,N}&=0 ~~~ (N>2)\\
C_{3,N}&=0 ~~~ (\text{for all~} N) \\
C_{4,N}&=0 ~~~ (N>4)\\
C_{6,N}&=0 ~~~ (N>6)\,,
\end{align}
we can match the BSQED and relativistic Ritz results, term by term in a double expansion in small $Z\a$ and $n^{-1}$. Suppressing angular momentum indices for clarity, we find
\begin{eqnarray}\label{delta_naught_estimate}
\de_{(0)}&=&-\(Z \a\)^2 \(C_{4,3} + \(Z\a\) C_{5,3}+{\cal O}\(Z\a\)^2\)\\
&=& \f{\(Z\a\)^2}{2j+1}+{\cal O}\(Z\a\)^3\notag\,,
\end{eqnarray}
which is consistent with the fit values from Tables \ref{table:fit_parameters_S_f0}, \ref{table:fit_parameters_P_f0}, and \ref{table:fit_parameters_D_f1}. Additional defect parameters are given in Appendix \ref{QED_lessons}.

In fact, there are too few free parameters in the relativistic Ritz model to account for all terms in \eqref{E_QED_asymp_form} unless the following relations hold true:
\begin{eqnarray}
C_{2,2}&=&-\f{1}{2}\\
C_{4,4}&=&\f{1}{8}\(3-\m+\m^2\)\label{C44}\\
C_{5,N}&=&0~~~~~~~~~~~~~~~~~~~~~~~\text{(even $N$)}\label{C5N}\\
C_{6,4}&=&-\f{3}{2}C_{4,3}^2\label{C64}\\
C_{6,6}&=&\f{1}{16}\(-5+3\m-4\m^2+2\m^3-\m^4\)\label{C66}\\
C_{7,4}&=&-3 C_{4,3} C_{5,3}\label{C74}\\
C_{7,6}&=&-5 C_{4,3}C_{5,5}\label{C76}\\
C_{7,8}&=&-7 C_{4,3}C_{5,7}\label{C78}\\
C_{7,10}&=&-9 C_{4,3}C_{5,9}\label{C710}
\end{eqnarray}
\begin{eqnarray}
C_{8,4}&=&-\f{3}{2}\(C_{5,3}^2 +2C_{4,3}C_{6,3}\)\\
C_{8,6}&=&-\f{5}{4}\(3-\m+\m^2\)C_{4,3}^2 \notag\\
&&~~~~-5 \(C_{5,3}C_{5,5}+C_{4,3}C_{6,5}\)\\
C_{8,8}&=&\f{1}{128}\Big(35 - 29\m+47\m^2-41\m^3 +33\m^4\notag\\
&&~~~~-15\m^5+5\m^6\Big)- \f{7}{2}C_{5,5}^2 - 7 C_{5,3} C_{5,7} \\
C_{8,10}&=&-9\(C_{5,5}C_{5,7}+C_{5,3}C_{5,9}\)\,,
\end{eqnarray}
and additional relations up to $A=10$ and $N=10$ are given in Appendix \ref{QED_lessons}.

It appears that all even-$N$ terms obey an asymptotic consistency relation\footnote{This even-odd correspondence is similar to that found by G.W.F. Drake in a non-relativistic context \cite{DRAKE199493}.} to odd-$N$ terms that is, remarkably, exact to all orders in the mass ratio $m_1/m_2$. Equation \eqref{C44} and \eqref{C66} may be readily checked in two special cases: (a) when $m_2\to \infty$ it follows that $C_{44}=3/8$ and $C_{66}=-5/16$; (b) when $m_2=m_1$  it follows that $C_{44}=11/32$ and $C_{66}=-69/256$. At intermediate mass ratios, treating $m_1/m_2$ as a small parameter, these results are consistent up to ${\cal O}\(m_1/m_2\)$ with results presented in Ref. \cite{Jentschura:2008zz}, whose results are based off of Ref. \cite{IEIDES200163}. All terms up to ${\cal O}(\a^6)$ are confirmed, including the non-linear relation \eqref{C64}. Equations \eqref{C5N} and \eqref{C74} through \eqref{C710} have implications for the Bethe logarithm which we considered in the following section.

\sec{Implications for Bethe Logarthims}\label{Bethe_Log_Discussion}

Beyond the leading order Bohr and fine-structure correction terms, proportional to $\(Z\a\)^2$ and $\(Z\a\)^4$, respectively, there are quantum field-theoretic effects predicted within BSQED to contribute to the energy at order $\a\(Z\a\)^4$ or, equivalently, $Z^{-1}\(Z\a\)^5$. These terms arise from self energy, vacuum polarization, and anomalous magnetic-moment corrections \cite{Drake:1990zz}. 

At second order in perturbation theory, the leading one-loop self-energy corrections include terms proportional to
\begin{equation}\label{terms_w_BetheLog}
\frac{m_1 \mu^3}{n^3}\f{\(Z\a\)^5}{Z}\ln{k_0(n,\ell)}\,.
\end{equation}
where $\ln{k_0(n,\ell)}$ is the called the nonrelativistic Bethe logarithm -- it is nonrelativistic in the sense that it is computed in perturbation theory using the unperturbed nonrelativistic wavefunctions of the Schroedinger-Coulomb equation. Roughly speaking, $k_0(n,\ell)$ is a normalized weighted average of excitation energies between the bound state with quantum numbers $n$ and $\ell$ and all other intermediate states (bound and continuum) that are reached during the emission and absorption of the virtual photon \cite{BetheSalpeter}. 
The Bethe logarithm is a pure number that has no known method of analytic calculation, so it must be computed numerically for a given $n$ and $\ell$. More than 20,000 values have been have been computed, up to $n=200$ and $\ell=199$, by the authors of \cite{PhysRevA.72.012110}, although conceivably values could be needed to even higher values of $n$. Investigations have demonstrated that it asymptotes to an $l$-dependent constant as $n\to\infty$ and may be fit with a series expansion in inverse powers of $n$:
\begin{equation}\label{general_Bethe_ansatz}
\ln{k_0(n,\ell)}\sim \ln{k_0(\infty,\ell)} + \f{\b_\ell^{(1)}}{n}+ \f{\b_\ell^{(2)}}{n^2} + \f{\b_\ell^{(3)}}{n^3} +\dots
\end{equation}
However, as they appear in terms like \eqref{terms_w_BetheLog}, Bethe logarithms are predicted by equation \eqref{C5N} to admit an asymptotic expansion that excludes inverse \emph{odd} powers of $n$. Using the numerical results of \cite{PhysRevA.72.012110}  for $n=190$ through $n=200$, some of which are reproduced in Table \ref{table_of_Bethe_logs}, we can fit for the parameters indicated in the general fit equation \eqref{general_Bethe_ansatz} as well as in the even-only fit in which the restriction $\b_\ell^{(i=\text{odd})}=0$ is made.

\begin{table}[H]
\centering
\caption{Numerically computed values of the Bethe logarithm from \cite{PhysRevA.72.012110}. }
\begin{tabular}{c | c c c}  \hline\hline
$\ln{k_0}(n,\ell)$ & $\ell=0$ & $\ell=1$ & $\ell=2$  \\ \hline
$n=190$ & $2.72266958$ &  $-0.0490489444$ &  $-0.00993 712 588$\\  
$n=191$ & $2.72266942$ &  $-0.0490490025$ &  $-0.00993 716 023$\\  
$n=192$ & $2.72266927$ &  $-0.0490490596$ &  $-0.00993 719 406$\\  
$n=193$ & $2.72266911$ &  $-0.0490491159$ &  $-0.00993 722 737$\\  
$n=194$ & $2.72266896$ &  $-0.0490491713$ &  $-0.00993 726 017$\\  
$n=195$ & $2.72266881$ &  $-0.0490492258$ &  $-0.00993 729 247$\\  
$n=196$ & $2.72266867$ &  $-0.0490492796$ &  $-0.00993 732 428$\\  
$n=197$ & $2.72266852$ &  $-0.0490493325$ &  $-0.00993 735 562$\\  
$n=198$ & $2.72266838$ &  $-0.0490493846$ &  $-0.00993 738 649$\\  
$n=199$ & $2.72266824$ &  $-0.0490494360$ &  $-0.00993 741 690$\\  
$n=200$ & $2.72266810$ &  $-0.0490494865$ &  $-0.00993 744 687$\\  \hline\hline
\end{tabular}
\label{table_of_Bethe_logs}
\end{table}

In Tables \ref{table_of_fit_parameters_l=0}, \ref{table_of_fit_parameters_l=1}, and \ref{table_of_fit_parameters_l=2} the best-fit parameters are shown as well as a comparison of the second-order Akaike Information Criterion (AICc) of each fit, made using  \emph{Mathematica} to the 11 entries of each column. Ref \cite{anderson2004model} contains an excellent discussion of the use of AICc for comparing the goodness-of-fit of various models to data. This criterion is defined by
\begin{equation}
\text{AICc} = -2 \log{\cal L(\hat{\theta})} + 2K\(\frac{n}{n-K-1}\)\,,
\end{equation}
where ${\cal L(\hat{\theta})}$ is the maximum likelihood for the set of best-fit model parameters ($\hat{\theta}$), $n$ is the sample size, and $K$ is the number of model parameters.   The ``best" model has the minimum value of AICc. AICc, as opposed to AIC, is used according to the rule of thumb that the former should be used whenever $n/K<40$ \cite{anderson2004model}.

\begin{table}[h]
\centering
\caption{\raggedright Numerical fits of the $\ell=0$ Bethe logarithm using data found in Table \ref{table_of_Bethe_logs}.}
\begin{tabular}{c c c}  \hline\hline
~ & General fit& Even-only fit \\ \hline
$\ln{k_0(\infty,0)}$ & $2.709(21)$ &  $2.72265425(85)$\\  
$\b_0^{(1)}$ & $10(16)$ &  --\\  
$\b_0^{(2)}$ & $-3.0(4.7)\times10^3$ &  $0.560(65)$\\  
$\b_0^{(3)}$ & $3.9(6.1)\times10^5$ &  --\\  
$\b_0^{(4)}$ & $-1.9(3.0)\times10^7$ &  $-232(1230)$\\  \hline
AICc & $-373.2$ &  $-391.8$\\  \hline\hline
\end{tabular}
\label{table_of_fit_parameters_l=0}
\end{table}

\begin{table}[h]
\caption{\raggedright Numerical fits of the $\ell=1$ Bethe logarithm as in Table \ref{table_of_fit_parameters_l=0}.}
\centering
\begin{tabular}{c c c}  \hline\hline
~ & General fit& Even-only fit\\ \hline
$\ln{k_0(\infty,1)}$ & $-0.04874(23)$ &  $-0.049054521(10)$\\  
$\b_1^{(1)}$ & $-0.24(18)$ &  --\\  
$\b_1^{(2)}$ & $71(52)$ &  $0.20201(78)$\\  
$\b_1^{(3)}$ & $-9.2(6.7)\times10^{3}$ &  --\\  
$\b_1^{(4)}$ & $4.5(3.3)\times10^5$ &  $-24(15)$\\  \hline
AICc & $-472.6$ &  $-489.0$\\  \hline\hline
\end{tabular}
\label{table_of_fit_parameters_l=1}
\end{table}

\begin{table}[h]
\centering
\caption{\raggedright Numerical fits of the $\ell=2$ Bethe logarithm as in Tables \ref{table_of_fit_parameters_l=0} and \ref{table_of_fit_parameters_l=1}.}
\begin{tabular}{c c c}  \hline\hline
~ & General fit& Even-only fit\\ \hline
$\ln{k_0(\infty,1)}$ & $-0.009980(16)$ &  $-0.009\,940\,4467(12)$\\  
$\b_1^{(1)}$ & $0.031(13)$ &  --\\  
$\b_1^{(2)}$ & $-8.9(3.8)$ &  $0.121026(93)$\\  
$\b_1^{(3)}$ & $1165(490)$ &  --\\  
$\b_1^{(4)}$ & $-5.7(2.4)\times10^4$ &  $-41.4(1.8)$\\  \hline
AICc & $-530.2$ &  $-535.9$\\  \hline\hline
\end{tabular}
\label{table_of_fit_parameters_l=2}
\end{table}

\begin{centering}

\begin{figure}[hp]
  \begin{center}
    \includegraphics[scale=.6]{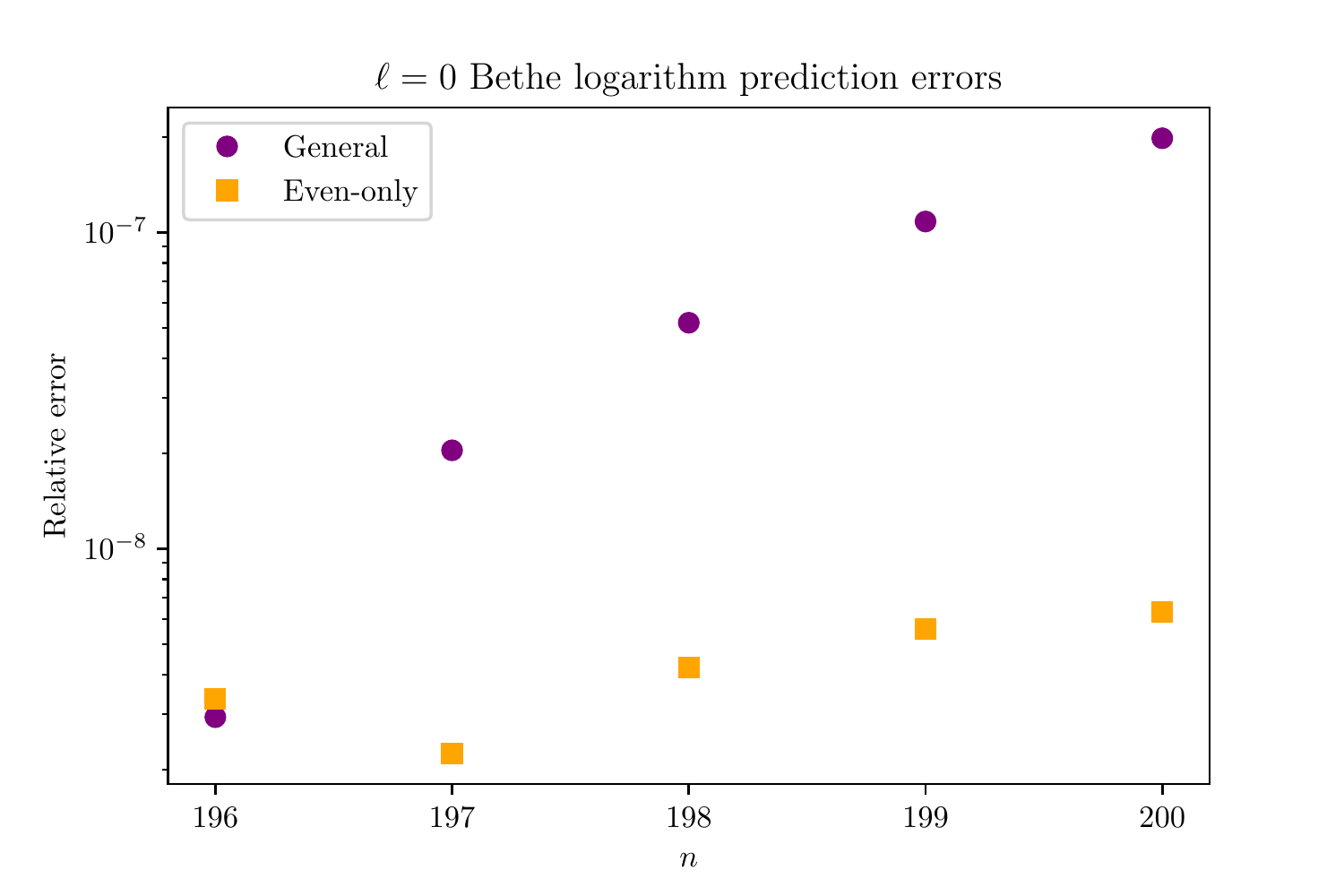}
  \end{center}
  \caption{\raggedright Relative errors on the $\ell=0$ Bethe logarithm predictions in the general and even-only fits using equation \eqref{general_Bethe_ansatz} to fit numerical data from Table \ref{table_of_Bethe_logs}, as described in the text.}%
  \label{BetheLogPred}
  \end{figure}

\end{centering}

Given the substantially lower value of the AICc\footnote{The relative evidence for one model compared to another depends exponentially on the AICc value. Given two models, the rule of thumb is that if one model has an AICc value higher by more than 10 then it has essentially no empirical support compared to the other model \cite{anderson2004model}.}, it is clear that a series of even inverse powers of $n$ is a superior model compared to a general series of inverse powers, \emph{at least for large values of $n$}, and should therefore provide a more accurate extrapolation to terms at even higher $n$. As an explicit demonstration, I have repeated the $\ell=0$ fit on only the first 6 entries ($n=190$ through $n=195$) of Table \ref{table_of_Bethe_logs} and made predictions for the remaining 5 entries ($n=196$ through $n=200$). The relative errors from the general and even-only fits are shown in Figure \ref{BetheLogPred}.

 It is also noteworthy that the asymptotic even-only fit parameters are in good agreement with the
numerical values $\ln{k_0(\infty,0)}=2.722\,654\,335$, $\ln{k_0(\infty,1)}=-0.049\,054\,544$, and $\ln{k_0(\infty,2)}=-0.009\,940\,457$ reported in \cite{PhysRevA.72.012110}. However, the even-only fits described here give inferior fits for smaller values of $n$, for example, the set of values for $n\leq 20$ computed and tabulated in Ref \cite{Drake:1990zz}. This is not surprising because the analyses discussed here use an asymptotic expansion around $n\to \infty$.

At ${\cal O}(\a^7)$, equations \eqref{C74} through \eqref{C710} indicate a relationship between the ${\cal O}(\a^5)$ Bethe logarithm and the higher order ``relativistic" Bethe logarithm, which is discussed, e.g., in \cite{PhysRevLett.90.163001} and \cite{LeBigot:2003zz}. However, this relationship has not been confirmed because the asymptotic behavior of this higher order Bethe logarithm is difficult to extrapolate; to my knowledge they have thus far only been computed for excited states from $n=2$ to $n=8$ \cite{LeBigot:2003zz}. Its computation to higher values of $n$ could confirm equations \eqref{C74} through \eqref{C710}. Alternatively, one line of research would be to assume their validity and pursue a more efficient method of computing the relativistic Bethe logarithm.

\sec{Discussion}\label{Sec:Discussion}

Here I have derived a relativistic long-distance effective theory of hydrogen-like atoms, dubbed the relativistic Ritz approach, and explored some of its consequences in the context of bound-state quantum electrodynamics (BSQED). The approach has been demonstrated to be superior to using canonical Rydberg-Ritz formula when applied to the hydrogen atom. This is undoubtedly also true for deuterium and other single-electron atoms. The approach should also be pursued for its application to the highly excited (Rydberg) states of more complex atoms. The alkali atoms, in particular, are used in contemporary studies of quantum information processing \cite{wu2021concise} and it is important to have accurate values of the (unperturbed) energy levels of these atoms \cite{mack2011measurement}. Plausibly, it may be possibly to modify this effective approach to be applied to in multi-electron or molecular systems.

Consistency relations within BSQED have been discovered that illuminate additional structure in the theory which may aid theorists in advancing of its predictive accuracy. In particular, new structure of the asymptotic form of the Bethe logarithm has been both predicted and confirmed through through analysis of numerical results. As a consequence, a more accurate method has been demonstrated for determining Bethe logarithms of arbitrarily large $n$ by extrapolating from currently known values at lower $n$. The existence of this structure within BSQED suggests that it may be possible to develop a more efficient method for some calculations that makes this structure manifest.  Conceivably, this could be achieved by  use of the unperturbed ($\de\to0$) relativistic solutions of equation \eqref{radial_Solution} to evaluate expectation values used in perturbation theory. Doing so consistently would require careful consideration that goes well beyond the scope of this article. 



This approach will be applied to atomic hydrogen and deuterium transition data, demonstrating its use for determining atomic ionization energies and the fine-structure constant directly from data. A straightforward application to data is complicated by the fact that some measured transitions occur between known hyperfine levels, while others are not hyperfine-resolved. This is work in preparation and will be described in a subsequent publication.

\begin{center}
{\bf Acknowledgements}
\end{center}
I am grateful for many constructive conversations at the conference PSAS`2022 in Warsaw, Poland. In particular, I would like to thank  Vojt\v ech Patk\'o\v s, Krzysztof Pachucki, Jean-Phillippe Karr, and Gordon Drake. 

\appendix

%
%
%
%

\sec{Relation to earlier work of Crater et al.}\label{Crater_et_al}

 In Ref.  \cite{Crater:1992xp}, the variable
\begin{equation}
w=E + m_1 + m_2
\end{equation}
is used for the center of mass energy. Using their definitions
\begin{equation}
\epsilon_w\equiv\frac{w^2-m_1^2-m_2^2}{2w}
\end{equation}
and
\begin{equation}
b^2(w)\equiv \f{w^4-2w^2\(m_1^2+m_2^2\)+\(m_1^2-m_2^2\)^2}{4w^2}\,,
\end{equation}
the ``weak-potential" eigenvalues to the Schrodinger form of the 2-body Dirac equation are found in Ref \cite{Crater:1992xp} to obey
\begin{equation}
\f{-b^2(w)}{\(\epsilon_w \alpha\)^2}=\frac{1}{n^2}\,,
\end{equation}
which, upon algebraic manipulation, yields \eqref{general_E_sol} when $n_\star\to n$\footnote{During manuscript preparation I also became aware of the unpublished article by J.H. Connell \cite{Connell:2012jp} in which a very similar ``two-body Sommerfeld" formula appears nearly identical to \eqref{general_E_sol}. However, the markedly different derivation attempted in that work and a subsequent unpublished article \cite{Connell:2017dra} was incomplete by Prof. Connell's admission. Furthermore, as in Ref. \cite{Crater:1992xp}, in that approach no connection is made to quantum defect theory or long-distance effective theories, in general.}.

\sec{More details from QED matching}\label{QED_lessons}


Letting
\begin{equation}
f_1(\mu)=3-\mu+\mu^2
\end{equation}
and
\begin{equation}
f_2(\mu)=-3+3\mu-2\mu^2-2\mu^3+\mu^4\,,
\end{equation}
it may be verified that the matching described in Section \ref{QED_insights} results in the following values for the defect parameters
\begin{widetext}
\begin{equation}
\de_{(0)}=-\(Z \a\)^2 \(C_{4,3} + \(Z\a\) C_{5,3}+ \(Z\a\)^2 C_{6,3}+\(Z\a\)^3 C_{7,3} \)+{\cal O}\(Z\a\)^6\,,
\end{equation}
\begin{equation}
\de_{(2)}=\f{\(Z\a\)^2}{2}f_1(\mu)\de_{(0)} -\(Z\a\)^3\(C_{5,5}+\(Z\a\) C_{6,5}+\(Z\a\)^2C_{7,5}+{\cal O}\(Z\a\)^3\) 
\end{equation}
\begin{equation}
\de_{(4)}= -\(Z\a\)^3\(C_{5,7}+\(Z\a\)^2\(C_{7,7}+\f{1}{2}f_1(\m)C_{5,5} \)\)+{\cal O}\(Z\a\)^6\,,
\end{equation}
and
\begin{equation}
\de_{(6)}= -\(Z\a\)^3\(C_{5,9} + \(Z\a\)^2\(C_{7,9}+\f{1}{2}f_1(\m)C_{5,7}\)\) +{\cal O}\(Z\a\)^6\,.
\end{equation}
The remaining consistency relations not listed in the main text are, up to order ${\cal O}\(Z\a\)^{10}$, 
\begin{eqnarray}
C_{9,4}&=&-3\(C_{6,3}C_{5,3}+C_{4,3}C_{7,3}\)\\
C_{9,6}&=&-5\(C_{6,3}C_{5,5}+ C_{4,3}C_{7,5}+C_{6,5}C_{5,3} +\f{1}{2}f_1(\mu)C_{4,3}C_{5,3}\)\\
C_{9,8}&=&-7\(C_{6,3}C_{5,7}+ C_{4,3}C_{7,7}+C_{6,5}C_{5,5} +\f{1}{2}f_1(\mu)C_{4,3}C_{5,5}\)\\
C_{9,10}&=&-9\(C_{6,3}C_{5,9}+ C_{4,3}C_{7,9}+C_{6,5}C_{5,7} +\f{1}{2}f_1(\mu)C_{4,3}C_{5,7}\)
\end{eqnarray}
\begin{eqnarray}
C_{10,4}&=&-3\(C_{4,3}C_{8,3}+C_{5,3}C_{7,3}+\f{1}{2}C_{6,3}^2\)\\
C_{10,6}&=&-5\(C_{4,3}C_{8,5}+C_{5,3}C_{7,5}+C_{6,3}C_{6,5}+C_{7,3}C_{5,5}-\f{3}{2}C_{4,3}^4+\f{1}{4}f_1(\mu)C_{5,3}^2+\f{1}{2}f_1(\mu)C_{4,3}C_{6,3}\)\\
C_{10,8}&=&-7\(C_{4,3}C_{8,7}+C_{5,3}C_{7,7} + \f{1}{2}C_{6,5}^2+C_{7,3}C_{5,7}+C_{5,5}C_{7,5} +  \f{1}{2}f_1(\m)\(C_{5,3}C_{5,5}+C_{4,3}C_{6,5}\)\)\notag\\
&&~~~~~+\f{7}{16}C_{4,3}^2f_2(\m)\\
C_{10,10}&=&\f{1}{256}\Big(-63+65\m-122\m^2 +144\m^3-154\m^4+118\m^5-72\m^6 + 28\m^7 - 7\m^8\Big) \notag\\
&&~~~~~-\f{9}{4}f_1(\m)\Big(C_{5,5}^2+2C_{5,3}C_{5,7}\Big)  -9\Big( C_{5,9}C_{7,3} + C_{5,7}C_{7,5} + C_{5,5}C_{7,7} + C_{5,3}C_{7,9} + C_{4,3}C_{8,9}\Big)\,.
\end{eqnarray}
\end{widetext}

\bibliographystyle{apsrev}
\bibliography{RRitz}

\end{document}